\documentclass[showpacs,preprintnumbers,amsmath,amssymb]{revtex4}
\usepackage{amsmath,amssymb,graphics,epsfig,subfigure}
\usepackage{color}

\begin{document}
\renewcommand{\baselinestretch}{1.3}

\title{Testing the nature of Gauss-Bonnet gravity by four-dimensional rotating black hole shadow}

\author{Shao-Wen Wei \footnote{weishw@lzu.edu.cn}}
\author{Yu-Xiao Liu \footnote{liuyx@lzu.edu.cn, corresponding author.}}

\affiliation{Institute of Theoretical Physics $\&$ Research Center of Gravitation, Lanzhou University, Lanzhou 730000, China,\\
 Lanzhou Center for Theoretical Physics $\&$ Key Laboratory of Theoretical Physics of Gansu Province, Lanzhou University, Lanzhou 730000,  China,\\
 Joint Research Center for Physics, Lanzhou University and Qinghai Normal University, Lanzhou 730000 and Xining 810000, China}

\begin{abstract}
The recent discovery of the novel four-dimensional static and spherically symmetric Gauss-Bonnet black hole provides a promising bed to test Gauss-Bonnet gravity by using astronomical observations [Phys. Rev. Lett. 124, 081301 (2020)]. In this paper, we first obtain the rotating Gauss-Bonnet black hole solution by using the Newman-Janis algorithm, and then study the shadow cast by the nonrotating and rotating candidate Gauss-Bonnet black holes. The result indicates that positive metric parameter $\alpha$ shrinks the shadow, while negative one enlarges it. Meanwhile, both the distortion and ratio of two diameters of the shadow are found to increase with the metric parameter for certain spin. Comparing with the Kerr black hole, the shadow gets more distorted for $\alpha$, and less distorted for negative one. Furthermore, we calculate the angular diameter of the shadow by making use of the observation of M87*. The result indicates that negative metric parameter $\alpha$ in (-4.5, 0) is more favored. Since the negative energy appears for negative $\alpha$, our results extends the study of Gauss-Bonnet gravity. We believe further study on the four-dimensional rotating black hole may shed new light on Gauss-Bonnet gravity.
\end{abstract}



\maketitle

\section{Introduction}

A couple of years ago, Event Horizon Telescope (EHT) Collaboration released the first image of the supermassive black hole located at the center of M87 galaxy \cite{Akiyama1,Akiyama2,Akiyama3,Akiyama4,Akiyama5,Akiyama6}. This fruitful outcome shows a fine structure near a black hole horizon and opens up a new window to test the strong gravity regime. The result indicates that the black hole shadow has a diameter 42$\pm$3 $\mu$as. Modeled M87* with a Kerr geometry, the observations were found to be well consistent with the prediction of general relativity. However, there still exists a living space for some modified gravities due to the finite resolution. With future observations, like the Next Generation Very Large Array \cite{Hughes}, the Thirty Meter Telescope \cite{Sanders}, and the BlackHoleCam \cite{Goddi}, it will offer us a good opportunity to peek into the regime of strong gravity, and to distinguish different modified gravities. Meanwhile, more knowledge and information on modified gravities and quantum gravity will be greatly revealed.

EHT observation restimulates the study of the black hole shadow. As we know, when photons are emitted from its source, and then fly past a black hole, they will have three results---absorbed by the black hole, reflected by the black hole and then escaping to infinity, or surrounding the black hole one loop by one loop. These photons escaped from the black hole will illuminate the sky of an observer. However, these photons absorbed by the black hole will leave a dark zone for the observer. This dark zone is actually the black hole shadow.

As early as 1966, Synge studied the observed angular radius of a Schwarzschild black hole \cite{Synge}. Later, a formula describing the shadow size was given by Luminet \cite{Luminet}. Now it is generally known that the shadow shape of a spherically symmetric black hole is round. While it will be elongated, for a rotating black hole, in the direction of the rotating axis due to spacetime dragging effect \cite{Bardeen,Chandrasekhar}. The size and distortion of the shadow have a close relation with the spacetime geometry. Thus by constructing different observables, a number of papers concern how do these observables depend on the parameters of the black hole spacetime \cite{Hioki,Amarilla2,Johannsen,Ghasemi,Bambi,Amarilla,Stuchlik,Amarilla13,Nedkova,
Wei,Tsukamoto,Bambi3,Atamurotov,Mann,WWei,FAtamurotov2,AmirAhmedov,Songbai,Balendra,
Tsukamoto2,Jiliang,Rajibul,hou,Cunha,Cunha2,Cunha3,Tsupko,Perlick,Kocherlakota,
WangXu,WeiLiu,WeiLiu2,Younsi,cBambi,Akashaa,Abdujabbarov,Shaikh,Chenw,Freese,
Konoplyab,Vagnozzib,Zhub,Banerjeeb,Lua,Fengb,Renb,Guob,Zhuc}. There are other related works \cite{Toth,Davoudiasl,Bar,Tian3,Kumarg,Allahyari,Rummel,Kumarw,Narang} that aim to constrain the parameters of the black holes or other compact objects from astronomical observations of M87*. Moreover, it is also expected to cast deep insight into modified gravities.

In the past few decades, different modified gravity theories were proposed with the attempt to solve the fundamental questions, such as the quantum gravity and singularity problem. Among them, Gauss-Bonnet (GB) gravity including higher curvature corrections is one of the most promising approaches. It is well known for a long time that there exist different static and spherically symmetric black hole solutions from general relativity in $d$-dimensional spacetime with $d>4$. While no different black hole solution exists in four dimensions due to that the GB term is a total derivative, and thus it has no contribution to the gravitational dynamics. Since the dimension of the observed spacetime is four, it is extremely hard to test the nature of GB gravity through astronomical observations.

In Refs. \cite{Tomozawa,Cognola}, the authors considered the GB gravity in four dimensions by rescaling the GB coupling parameter $\alpha\rightarrow\frac{\alpha}{d-4}$. Very recently, Glavan and Lin \cite{Glavan} reconsidered this issue and proposed a general covariant modified gravity in four dimensions, in which only the massless graviton propagates. It can also bypass the Lovelock's theorem and avoid Ostrogradsky instability. Taking the dimension number $d\rightarrow4$, the GB term shows a nontrivial contribution to the gravitational dynamics. And then a nontrivial and novel four-dimensional static and spherically symmetric black hole solution was discovered. Such black hole, in particular, offers us a promising bed to test the nature of GB gravity. Subsequently, the quasinormal modes of scalar, electromagnetic and gravitational perturbations were calculated in \cite{Zinhailo}, where the results show that varying with the GB coupling parameter, the damping rate is more sensitive characteristic than the real part. Dynamical eikonal instability occurs for larger GB coupling parameter. The shadow cast by the spherically symmetric black hole was examined in Refs. \cite{Zinhailo,Guoli}. The shadow size exhibits a close relation with the coupling parameter. In addition, all the radii of the innermost stable circular orbit, black hole horizon, and the photon sphere are decreasing functions of the GB coupling parameter \cite{Guoli}. This black hole solution was generalized to the charged case \cite{Fernandes}. The cosmological and black hole solutions arising from the gravity was also discussed in Ref. \cite{Casalino}.

The singularity problem was also raised in this GB gravity in Ref. \cite{Glavan}. Glavan and Lin suggested that, although there is a singularity at $r$=0, the gravitational force was repulsive at a short distance, which will lead to that an infalling particle never reaches the $r$=0 point, and thus this theory is practically free from the singularity problem. However, following the study of the geodesics of a free-falling massive particle, it was shown that if an infalling particle starts at rest, no matter what its initial position is, it will reach  the singularity with zero velocity \cite{ArrecheaDelhom0,ArrecheaDelhom}.

Although this novel gravity admits a nontrivial contribution in four dimensions, there are works concerning about whether this gravity makes sense in general. Different multiple pathologies were pointed out in Refs.~\cite{LuPang,Ai,Gurses,Mahapatra,Hennigarq,Tian,Arrechea,Aoki,LiuLiuLiu,Hennigar5}. For example in Ref. \cite{Gurses}, the authors showed that such gravity does not admit a description in terms of a covariantly-conserved rank-2 tensor in four dimensions. Other references argued that the dimensional regularization procedure is ill-defined in a general spacetime. However, some regularized 4D EGB theories were presented in Refs.~\cite{Fernandes2,Hennigar1}. The regularization procedures include introducing an extra GB term constructed by a conformal metric~\cite{Fernandes2,Hennigar1} and compactification of the $D$-dimensional EGB theory~\cite{LuPang}. The resulting gravity has an extra scalar degree of freedom which helps us to obtain the field equations in a 4D version. Generally, it is believed that the regularization can be applied to the maximally symmetric or spherical symmetric space in 4D spacetimes, while may be problematic for the case of the axially symmetric case \cite{Hennigar1}. This approach suggested that the GB gravity belongs to the family of Horndeski gravity. Following Ref.~\cite{Hennigarq}, the treatment was also found to be applicable in a spacetime with low symmetry, for example the cylindrically symmetric spacetime \cite{LiuLiuLiu}. In the three dimensional case, the authors obtained two distinct but similar versions of the theory and the black hole solutions \cite{Hennigar5}. The results indicate that the black hole obtained by taking $d\rightarrow$3 is not the solution of the lower dimensional GB gravity except for particular constraints on the parameters. Nevertheless, it is still worth to investigate the novel properties for the rotating solutions in four dimensions for this gravity.

It is generally believed that almost all the astronomical black holes have spin. So it is worth to study the particular properties for the rotating black hole, which will provide us opportunities to test the nature of GB gravity by using astronomical observations, especially the observations of M87* by the EHT Collaboration. In this paper, we mainly focus on studying the shadow cast by the GB black hole, and constraining the metric parameter $\alpha$. So we first generalize the black hole to a rotating one by using the Newman-Janis (NJ) algorithm~\cite{Janis}. Although it was found in Ref.~\cite{Hansen} that the NJ trick is not generally applicable in higher curvature theories in a vacuum, the solution might describe a black hole with matter fields such as the fluids~\cite{Azregainou}. For this generalized GB gravity, since there are no the explicit field equations, the NJ algorithm provides us with a reasonable first step towards the construction of a rotating black hole solution. This will also help us further understanding the nature of the gravity. On the other hand, this black hole solution can be viewed as a modified Kerr black hole. In particular, it is worth to point out that the nonrotating black hole solution is the same as Refs. \cite{caicao,cai3} in the conformal anomaly gravity. So this solution can also be treated as a generalization in modified gravities. It provides us with a potential test of the supermassive black holes located at the center of galaxies in other relevant gravity theories. Then, by modeling M87* with this rotating black hole, we constrain the metric parameter $\alpha$ via the observations of EHT Collaboration. The result reveals that negative parameter is more favored. Another study on the shadow in Eeinstein-dilaton-Gauss-Bonnet black holes can be found in Ref. \cite{Cunhah}.

The paper is organized as follows. In Sec. \ref{foursbh}, we first show the nonrotating GB black hole solution, and then extend it to its rotating counterpart by the NJ algorithm. Different regions of the spacetime in the parameter space are also displayed. Null geodesics and circular photon orbit are given in Sec. \ref{shadows}. In Sec. \ref{observa}, the shadow shapes are exhibited. Basing on them, we construct several observables and obtain their behaviors with the spin and the metric parameter $\alpha$, which provides us the preliminary nature on the four-dimensional GB black hole. Comparing with the Kerr black hole, we observe that positive $\alpha$ shrinks the shadow, while negative one enlarges it. Then, after obtaining these results, we constrain the metric parameter by calculating the angular diameter of the shadow via the observation of M87* in Sec. \ref{M87}. Finally, the conclusions and discussions are presented in Sec. \ref{Conclusions}.

\section{Four-dimensional Gauss-Bonnet black hole}
\label{foursbh}

In this section, we will focus on the properties of the four-dimensional nonrotating and rotating black holes.

\subsection{Non-rotating black hole}
\label{nospin}

In GB gravity, it is well known that there are the static and spherically symmetric black hole solutions in a spacetime with $d\geq5$, for examples see Refs. \cite{Boulware,Wiltshire,caii,Nojiria,Cvetica}. However, in four-dimensional spacetime, the GB term is a total derivative, and thus it has no contribution to the gravitational dynamics. Until recently, a four-dimensional nontrivial black hole was discovered by Glavan and Lin \cite{Glavan}. They rescaled the GB coupling parameter $\alpha\rightarrow\alpha/(d-4)$, then took the limit $d\rightarrow4$, and finally found a static and spherically symmetric black hole solution \cite{Glavan}
\begin{eqnarray}
 ds^2&=&f(r)dt^2-\frac{dr^2}{f(r)}-r^2(d\theta^2+\sin^2\theta d\phi^2),\label{GBme}\\
 f(r)&=&1+\frac{r^2}{2\alpha}\left(1-\sqrt{1+\frac{8\alpha M}{r^3}}\right).\label{meme}
\end{eqnarray}
This black hole solution exactly coincides with that obtained in a gravity with conformal anomaly \cite{caicao,cai3}. Here $M$ is the black hole mass. As we know, GB gravity arises from low energy limit from heterotic string theory. The GB coupling parameter $\alpha$ has dimensions of the square of length and plays the role of the inverse string tension. The black hole horizons can be obtained by solving $f(r)=0$, which gives
\begin{equation}
 r_\pm=M\pm\sqrt{M^2-\alpha}.
\end{equation}
For positive $\alpha$, we have two horizons for $\alpha/M^2<1$, one degenerate horizon corresponding to an extremal black hole for $\alpha/M^2=1$, while no horizon for $\alpha/M^2>1$.

This solution can be obtained by solving the following action
\begin{eqnarray}
 S&=&\int\sqrt{-g}d^4x\left(R+\alpha L_{GB}\right),\\
 L_{GB}&=&R_{\mu\nu\rho\sigma}R^{\mu\nu\rho\sigma}-4R_{\mu\nu}R^{\mu\nu}+R^2,
\end{eqnarray}
where we have taken $16\pi G=1$. This solution can bypass the Lovelock's theorem and avoid Ostrogradsky instability. When $r\rightarrow0$ and $r\rightarrow\infty$, we have
\begin{eqnarray}
 f(r\rightarrow0)&=&1-\sqrt{\frac{2M}{\alpha}}r^{\frac{1}{2}}+\frac{1}{2\alpha}r^2
 +\mathcal{O}\left(r^{\frac{7}{2}}\right),\label{agd}\\
 f(r\rightarrow\infty)&=&1-\frac{2M}{r}+\frac{4\alpha M^2}{r^4}+\mathcal{O}\left(\frac{1}{r^7}\right).
\end{eqnarray}
So, this metric is asymptotic flat. From (\ref{meme}), we observe that at the short radial distances, the metric function will be complex for negative $\alpha$ and thus no real solution exists. This can also be found from (\ref{agd}), where the second part becomes imaginary near $r$=0 for negative $\alpha$. A detailed analysis shows that no real solution exists at short radial distances when
\begin{equation}
 r< r_0=2(-\alpha M)^{\frac{1}{3}}.
\end{equation}
However in the parameter range $-8\leq\alpha/M^2<0$, the singular short radial distances are always hidden behind the outer horizon $r_+$ \cite{Guoli}, which therefore provides a well behaved external solution. On the other hand, although $\alpha$ acts as the inverse string tension and should be positive, the solution (\ref{meme}) allows the existence of a negative $\alpha$. This might extend the conventional GB gravity. So it is interesting to examine the properties of this extended GB gravity with a negative $\alpha$. We expect some interesting properties could be revealed. Therefore, we will consider the black hole solution in the region $-8\leq\alpha/M^2<1$ in this paper.

\subsection{Rotating black hole}
\label{spin}

In this subsection, we would like to adopt the NJ algorithm \cite{Janis} to generate the rotating CGB black hole solution from (\ref{GBme}) by following the approach \cite{Azregainou}.

First, we introduce the Eddington-Finkelstein coordinates ($u$, $r$, $\theta$, $\phi$) with
\begin{equation}
 du=dt-\frac{dr}{f(r)}.
\end{equation}
Then the metric of the nonrotating GB black hole becomes
\begin{equation}
 ds^2=f(r)du^2+2dudr-r^2d\theta^2-r^2\sin^2\theta d\phi^2.\label{sur}
\end{equation}
Further, the metric can be expressed as
\begin{equation}
 g^{ab}=l^am^b+l^bn^a-m^a\bar{m}^b-m^b\bar{m}^a,
\end{equation}
with the null tetrads given by \cite{Azregainou}
\begin{eqnarray}
 l^a&=&\delta^a_r,\\
 n^a&=&\delta^a_\mu-\frac{f(r)}{2}\delta^a_r,\\
 m^a&=&\frac{1}{\sqrt{2}r}\left(\delta^a_\theta+\frac{i}{\sin\theta}\delta^a_\phi\right).
\end{eqnarray}
It is easy to find that these null tetrads have the following relations:
\begin{eqnarray}
 l^al_a=n^an_a=m^am_a=\bar{m}^a\bar{m}_a=0,\\
 l^am_a=l^a\bar{m}_a=n^am_a=n^a\bar{m}_a=0,\\
 l^an_a=-m^a\bar{m}_a=1.
\end{eqnarray}
Now, we perform the complex coordinate transformations in ($u$, $r$)-plane following the NJ algorithm
\begin{eqnarray}
 &&u'\rightarrow u-ia\cos\theta,\nonumber\\
 &&r'\rightarrow r+ia\cos\theta,\label{rp}
\end{eqnarray}
where $a$ is a spin parameter of the black hole.

\begin{figure}
\center{
\includegraphics[width=9cm]{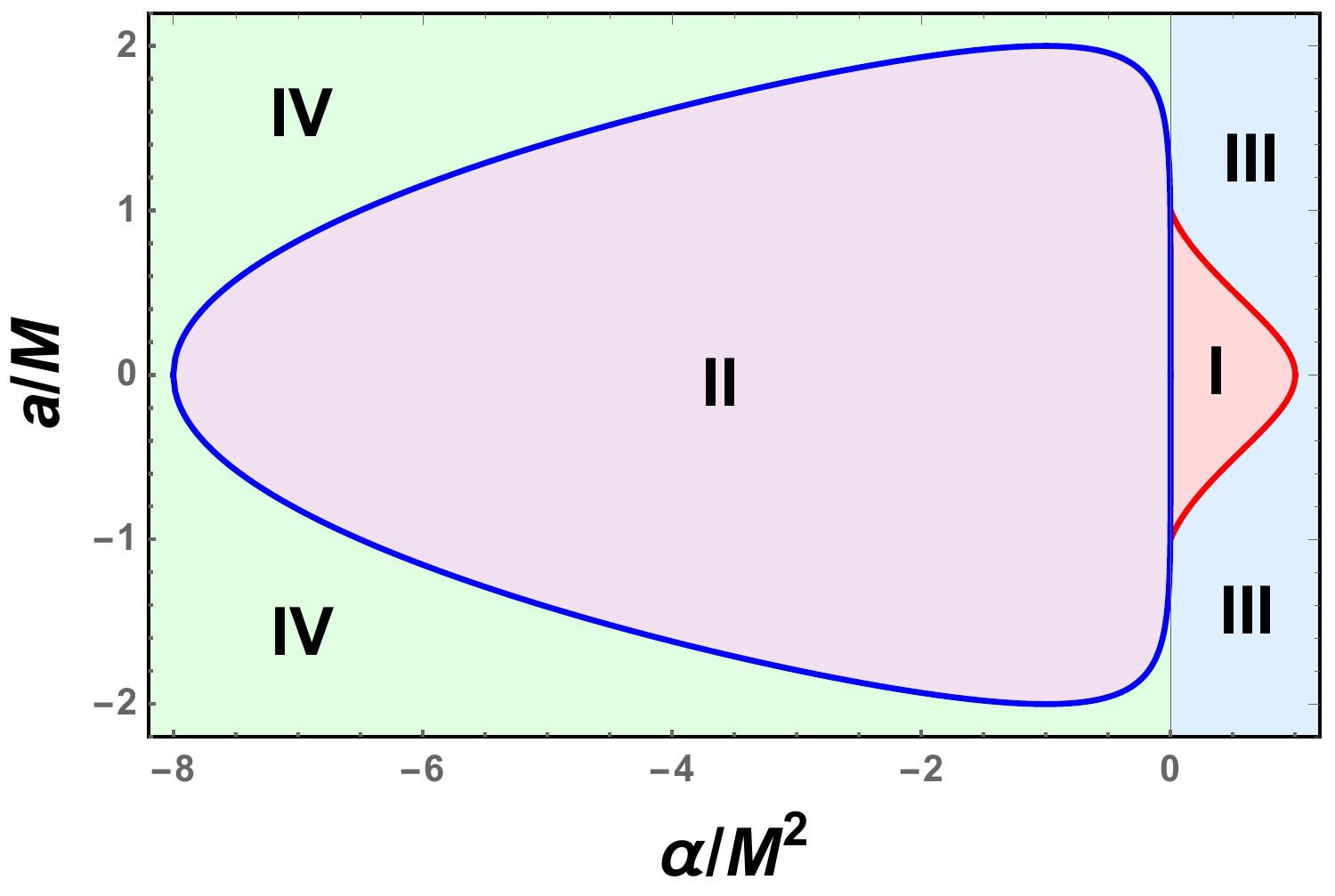}}
\caption{Regions I and II are black holes with two and one horizon, respectively. Region III is for the naked singularity. In region IV, the spacetime does not have horizon and real solution at the short radial distances.}\label{ppPT}
\end{figure}

Next step is complexifying the radial coordinate $r$ in the NJ algorithm. However it is not necessary. As shown by \cite{Azregainou}, this complexifying process can be dropped by considering that $\delta^\mu_\nu$ transforms as a vector under (\ref{rp}). At the same time the metric functions of (\ref{sur}) transform to new undetermined ones
\begin{eqnarray}
 &&f(r)\rightarrow F(r, a, \theta),\\
 &&r^2\rightarrow H(r, a, \theta).
\end{eqnarray}
After this transformation, the null tetrads become
\begin{eqnarray}
 l^a&=&\delta^a_r,\\
 n^a&=&\delta^a_\mu-\frac{F}{2}\delta^a_r,\\
 m^a&=&\frac{1}{\sqrt{2H}}\left((\delta^a_\mu-\delta^a_r)ia\sin\theta+\delta^a_\theta
 +\frac{i}{\sin\theta}\delta^a_\phi\right).
\end{eqnarray}
Making use of the new null tetrads, the rotating metric in the Eddington-Finkelstein coordinates is given by
\begin{eqnarray}
 ds^2&=&Fdu^2+2dudr+2a\sin^2\theta(1-F)dud\phi-2a\sin^2\theta drd\phi\nonumber\\
 &&-Hd\theta^2-\sin^2\theta\left(H+a^2\sin^2\theta(1-F)\right)d\phi^2.
\end{eqnarray}
Bringing this coordinates back to the Boyer-Lindquist ones, we can obtain the rotating GB black hole. In order to achieve it, we introduce a global coordinate transformation
\begin{eqnarray}
 du&=&dt+\lambda(r)dr,\\
 d\phi&=&d\phi'+\chi(r)dr,
\end{eqnarray}
with \cite{Azregainou}
\begin{eqnarray}
 \lambda(r)&=&-\frac{a^2+r^2}{a^2+r^2f(r)},\\
 \chi(r)&=&-\frac{a}{a^2+r^2f(r)}.
\end{eqnarray}
At last, we choose
\begin{eqnarray}
 F&=&\frac{(r^2f(r)+a^2\cos^2\theta)}{H},\\
 H&=&r^2+a^2\cos^2\theta.
\end{eqnarray}
Then the rotating black hole metric reads
\begin{eqnarray}
 ds^2=-\frac{\Delta}{\rho^2}(dt-a\sin^2\theta d\phi)^2+\frac{\rho^2}{\Delta}dr^2+\rho^2d\theta^2
 +\frac{\sin^2\theta}{\rho^2}\left(adt-(r^2+a^2)d\phi\right)^2.\label{Romet}
\end{eqnarray}
Note that, we have changed the sign of the metric according to the convention. For the four-dimensional rotating candidate Gauss-Bonnet (CGB) black hole, the metric functions are
\begin{eqnarray}
 \rho^2&=&r^2+a^2\cos^2\theta,\\
 \Delta&=&r^2+a^2+\frac{r^4}{2\alpha}\left(1-\sqrt{1+\frac{8\alpha M}{r^3}}\right).\label{DDm}
\end{eqnarray}
It is worth to point out that a couple of days later after we submitted this paper to arXiv, there is another paper concerning the same rotating black hole case and its shadow \cite{Kumar2}. We have set $16\pi G$=1 in the action, so both the two black hole solutions are the same. After the NJ algorithm, the rotating CGB black hole solution can be viewed as a solution to the Einstein field equations when extra matter fields are included. The corresponding energy-momentum tensor can be calculated with~\cite{Azregainou}
\begin{eqnarray}
 T^{\mu\nu}=\epsilon e^{\mu}_{t}e^{\nu}_{t}+p_{r}e^{\mu}_{r}e^{\nu}_{r}
 +p_{\theta}e^{\mu}_{\theta}e^{\nu}_{\theta}
 +p_{\phi}e^{\mu}_{\phi}e^{\nu}_{\phi}.
\end{eqnarray}
Here $\epsilon$ and $e^{\mu}_{t,r,\theta,\phi}$ are the density and four-velocity vectors of the fluid. $p_{i}$ denotes the components of the pressure. $e_{t,r,\theta,\phi}$ are dual to the following four 1-forms
\begin{eqnarray}
 \omega^{t}=-\frac{\sqrt{\Delta}}{\rho}(dt-a\sin^2\theta d\phi),\quad
 \omega^{r}=\frac{\rho}{\sqrt{\Delta}}dr,\\
 \omega^{\theta}=\rho d\theta,\quad
 \omega^{\phi}=\frac{\sin\theta}{\rho}\left(adt-(r^2+a^2)d\phi\right).
\end{eqnarray}
On the other hand, as we know, in GR, the Kerr solution can be obtained from a Schwarzschild one by using the NJ algorithm, and both black holes are the vacuum solutions. However, when applying the NJ algorithm to the four-dimensional GB black hole, the rotating one is not a vacuum solution anymore. Furthermore, considering the gravitational field equations, one can calculate the total energy-momentum tensor. However, for this four-dimensional GB gravity, one could not do this since there are no explicit field equations. But this could be resolved according to Refs. \cite{LuPang,Fernandes2,Hennigar1}, where some regularized 4D EGB theories were constructed and the gravitational field equations were given. When the spin $a=0$, this metric will reduce to the static and spherically symmetric one (\ref{GBme}). On the other hand, at the small $\alpha$ limit, we have
\begin{equation}
 \Delta(\alpha\rightarrow0)=\Delta_{Kerr}+\frac{4M^2}{r^2}\alpha+\mathcal{O}(\alpha^2),
\end{equation}
where $\Delta_{Kerr}=r^2-2Mr+a^2$ for the Kerr black hole. It is clear that the metric parameter $\alpha$ modifies the Kerr solution. While at large distance, it reduces to the Kerr case. Therefore, for the reasons discussed above, it should be emphasized that the metric (\ref{Romet}) is a modification to the Kerr metric.
The horizons of the CGB black hole can be obtained by solving $\Delta=0$. For positive $0\leq\alpha/M^2\leq1$, it is similar with the Kerr case. There can be two horizons, one horizon, and no horizon. There exists a maximal value of spin $a$ for certain $\alpha$, beyond which only the naked singularity is presented. While for negative $\alpha$, only one positive real root of $\Delta=0$ is found, which indicates that there only exists one black hole horizon. Also, similar to the nonrotating black hole, the solution only exists for $r\geq r_0$. In order to guarantee the existence of the horizon and $r_+\geq r_0$, one should have $\Delta(r_0)\leq0$, which requires that the black hole spin satisfies the following relation
\begin{equation}
 a^2/M^2\leq 4(-\alpha/M^2)^{\frac{1}{3}}\left(2-(-\alpha/M^2)^{\frac{1}{3}}\right).
\end{equation}
At $\alpha=-M^2$, the spin approaches its maximum 2$M$. In Fig. \ref{ppPT}, we show the different regions of the spacetime in $a/M$-$\alpha/M^2$ plane. In regions I and II, they are black holes with two and one horizon, respectively. Region III is for the naked singularity.  Differently, in region IV, the spacetime has no horizon and has no real solution at the short radial distances. Therefore, we only consider the black hole regions I and II. This shall uncover some interesting properties of this four-dimensional CGB black hole.

Taking $\theta=\pi/2$, we obtain
\begin{equation}
 T_{00}=\frac{12(3a^2-2r+r^2)\alpha}{r^8}+\mathcal{O}(\alpha^2),
\end{equation}
expanding at small $\alpha$. Clearly, for the Kerr black hole with $\alpha=0$, one gets $T_{00}=0$, which indicates a vacuum solution. One interesting phenomenon suggests that negative $\alpha$ produces negative $T_{00}$ outside the black hole horizon. So negative energy appears for the case of $\alpha<0$, which implies that exotic matter is present. Therefore, it is much more interesting to extend the study to the negative $\alpha$ case.

Note that the metric (\ref{Romet}) admits timelike and angular Killing vectors, we can compute the corresponding Komar mass and angular momentum. For an asymptotically flat, stationary, axisymmetric spacetime (which is just the case considered here), the Komar mass and angular momentum are given by the following integrals over a spacelike two-sphere at infinity~\cite{Komar}:
\begin{eqnarray}
 M_{\text{Komar}} &=& \frac{1}{4\pi}  \oint_\infty \sqrt{\gamma^{(2)}} n_{\mu }\sigma_{\nu} \nabla^{\mu}k^{\nu} d^2x,\\
 J_{\text{Komar}} &=& -\frac{1}{8\pi}  \oint_\infty \sqrt{\gamma^{(2)}} n_{\mu }\sigma_{\nu} \nabla^{\mu}m^{\nu} d^2x.
\end{eqnarray}
where $k^{\mu}=\delta^{\mu}_t$ and $m^{\mu}=\delta^{\mu}_{\phi}$ are the Killing vectors associated with translations and rotations around the $z$-axis. The metric on the two-sphere at infinity is
$ \gamma_{ij}^{(2)} dx^i dx^j = r^2 d\theta^2 + r^2  \sin^2(\theta) d\phi^2 $. The two normal vectors $n_{\mu}=\delta_{\mu}^t$ and $\sigma_{\mu}=\delta_{\mu}^r$ satisfy the conditions $n_{\mu}n^{\mu}=-1$ and $\sigma_{\mu}\sigma^{\mu}=1$, respectively, and their non-vanishing components are given by
\begin{eqnarray}
 n_t &=& -\sqrt{ {g_{t\phi}^2} / {{g_{\phi\phi}}} - g_{tt} }
      = -1 + \frac{M}{r}+  \frac{M^2}{2 r^2}
        + \mathcal{O}\left(\frac{1}{r^3}\right) ,\\
 \sigma_r &=& \sqrt{ g_{rr} } =  1 + \frac{M}{r} + \frac{3 M^2 - a^2 \sin ^2(\theta )}{2 r^2}
          + \mathcal{O}\left(\frac{1}{r^3}\right) .
\end{eqnarray}
Then, we can obtain the following result
\begin{eqnarray}
 M_{\text{Komar}} &=& M,\\
 J_{\text{Komar}} &=& a M.
\end{eqnarray}

\section{Geodesics and circular photon orbits}
\label{shadows}

In this section, we would like to investigate the geodesics in the background of (\ref{Romet}). We will study the circular photon orbits of the black hole, which is a key quantity on examining the shadows cast by the four-dimensional CGB black hole.

The geodesics of a particle moving in this background can be obtained by solving the geodesic equation. Alternately, one can adopt the Hamilton-Jacobi approach. The Hamilton-Jacobi equation describing the particle is
\begin{equation}
 \frac{\partial S}{\partial\lambda}=-\frac{1}{2}g^{\mu\nu}\frac{\partial S}{\partial x^{\mu}}\frac{\partial S}{\partial x^{\nu}},\label{sequation}
\end{equation}
where $\lambda$ is the affine parameter. For this black hole background, there are two Killing fields $\partial_t$ and $\partial_\phi$, which give us two constants, the particle energy $E$ and orbital angular momentum $l$ along each geodesics
\begin{eqnarray}
 -E&=&g_{t\mu}\dot{x}^{\mu},\\
 l&=&g_{\phi\mu}\dot{x}^{\mu}.
\end{eqnarray}
Then the Jacobi action can be separated as
\begin{equation}
 S=\frac{1}{2}\mu^2\lambda-Et+l\phi+S_{r}(r)+S_\theta(\theta),\label{jaction}
\end{equation}
where $\mu^2$ is the rest mass of the particle. The functions $S_{r}(r)$ and $S_{\theta}(\theta)$, respectively, depends only on $r$ and $\theta$. Substituting the Jacobi action (\ref{jaction}) into the Hamilton-Jacobi equation (\ref{sequation}), one obtains
\begin{eqnarray}
 S_r(r)=\int^r\frac{\sqrt{\mathcal{R}(r)}}{\Delta(r)}dr,\\
 S_\theta(\theta)=\int^\theta\sqrt{\Theta(\theta)}d\theta,
\end{eqnarray}
with
\begin{eqnarray}
 \mathcal{R}&=&\Big[(r^2+a^2)E-al\Big]^2-\Delta\Big[\mu^2r^2+\mathcal{K}+(l-aE)^2\Big],\\
 \Theta&=&\mathcal{K}+\cos^2\theta\left(a^2(E^2-\mu^2)-l^2\sin^{-2}\theta\right),
\end{eqnarray}
where the Cater parameter $\mathcal{K}$ related to the Killing-Yano tensor field is another constant of the geodesics. Further combining $g_{\mu\nu}\dot{x}^{\mu}\dot{x}^{\nu}=-\mu^2$, we finally approach the following equation of motion for the particle in the background of a rotating CGB black hole
\begin{eqnarray}
 \rho^{2}\frac{dt}{d\lambda}&=&a(l-aE\sin^{2}\theta)
       +\frac{r^{2}+a^{2}}{\Delta}\Big(E(r^{2}+a^{2})-al\Big),\label{rhot}\\
 \rho^{2}\frac{dr}{d\lambda}&=&\pm\sqrt{\mathcal{R}},\label{Rad}\\
 \rho^{2}\frac{d\theta}{d\lambda}&=&\pm\sqrt{\Theta},\\
 \rho^{2}\frac{d\phi}{d\lambda}
     &=&(l\csc^{2}\theta-aE)+\frac{a}{\Delta}\Big(E(r^{2}+a^{2})-al\Big).\label{rhophi}
\end{eqnarray}
Now we focus on the circular photon orbits by analyzing the radial motion. We take $\mu^2=0$ for photons in the following discussions. The radial motion (\ref{Rad}) can be reexpressed as
\begin{equation}
 \left(\rho^{2}\frac{dr}{d\lambda}\right)^2+V_{eff}=0.
\end{equation}
The effective potential reads
\begin{equation}
 V_{eff}/E^2=-\Big[(r^2+a^2)-a \xi\Big]^2+\Delta\Big[\eta+(\xi-a)^2\Big],
\end{equation}
where $\xi=l/E$ and $\eta=\mathcal{K}/E^2$, and the function $\Delta$ is given in Eq. (\ref{DDm}). The unstable circular photon orbit satisfies the following conditions
\begin{equation}
 V_{eff}=0,\quad \frac{\partial V_{eff}}{\partial r}=0, \quad \frac{\partial^2 V_{eff}}{\partial r^2}<0. \label{qqq}
\end{equation}
The third condition ensures the orbit is unstable. Solving them, we obtain
\begin{eqnarray}
 \xi&=&\frac{\left(a^2+r^2\right) \Delta'-4 \Delta r}{a \Delta'},\label{xx1}\\
 \eta&=&\frac{r^2 \left(16 \Delta  \left(a^2-\Delta\right)-r^2 \Delta'^2+8
   \Delta  r \Delta'\right)}{a^2 \Delta'^2},\label{xx2}
\end{eqnarray}
where the prime denotes the derivative with respect to $r$. Inserting them into the unstable condition (\ref{qqq}), it requires that the radius of the unstable orbit satisfies
\begin{eqnarray}
 r+2\frac{\Delta}{\Delta'^2}(\Delta'-r\Delta'')>0,\label{ccdc}
\end{eqnarray}
where we have used $\Delta>0$ outside the horizon. It is easy to check that this condition holds for the Schwarzschild black hole, where $a=\alpha=0$, and the radius $r=3M$ of the photon sphere. This can also be numerically checked for the rotating CGB black hole.

\section{Black hole shadows and observables}
\label{observa}

In this section, we would like to study the shadows cast by the nonrotating and rotating CGB black holes. Before performing the study, it is worthwhile pointing out that in our case, all light sources are located at infinity and distributed uniformly in all directions. The observer is likewise located at infinity.

In order to describe the shadow under our assumption, one needs to introduce two celestial coordinates \cite{Bardeen}
\begin{eqnarray}
 X&=&\lim_{r\rightarrow \infty}
   \bigg(-r^{2}\sin\theta\frac{d\phi}{dr}
      \bigg|_{\theta\rightarrow \theta_{0}}\bigg)
     =-\xi\csc\theta_{0},\label{alpha}\\
 Y&=&\lim_{r\rightarrow \infty}
   \bigg(r^{2}\frac{d\theta}{dr}\bigg|_{\theta\rightarrow \theta_{0}}\bigg)
     =\pm\sqrt{\eta+a^{2}\cos^{2}\theta_{0}-\xi^{2}\cot^{2}\theta_{0}},\label{beta}
\end{eqnarray}
where the equations of motion (\ref{rhot})-(\ref{rhophi}) are used, and $\theta_0$ is the inclination angle of the observer. If the observer locates on the equatorial plane, these celestial coordinates simplify to
\begin{eqnarray}
 X&=&=-\xi\\
 Y&=&\pm\sqrt{\eta}.
\end{eqnarray}
The shadow can be obtained by producing a parametric plot in the $X$-$Y$ celestial plane consistent with Eqs. (\ref{xx1}) and (\ref{xx2}), where the parameter governing the plot is $r$. Such region is actually not illuminated by the photon sources. The boundary of the shadow can be determined by the radius of the circular photon orbits.

\subsection{Nonrotating black hole shadows}

Here we first consider the nonrotating black hole case, i.e., $a$=0, which is described by the metric (\ref{GBme}). On the other hand, because the spherical symmetry of the black hole, the inclination angle of the observer effectively equals to $\theta_0=\pi/2$.

For this nonrotating black hole with $a$=0, the unstable orbit is just the photon sphere of the black hole. Solving the first two conditions in (\ref{qqq}), one can easily obtain the photon sphere radius $r_{\text{ps}}$, which is \cite{Guoli}
\begin{equation}
 r_{\text{ps}}=2\sqrt{3}M\cos\left(\frac{1}{3}\arccos(-\frac{4\alpha}{3\sqrt{3}M^2})\right).\label{rs}
\end{equation}
When $\alpha=0$, this gives $r_{\text{ps}}=3M$, which is just the result of the Schwarzschild black hole case. When $\alpha$ takes its upper and lower bounds, i.e., -8$M^2$ and $M^2$, we have $r_{\text{ps}}=4.7428M$ and $2.3723M$, respectively. The relation (\ref{rs}) shows that $r_{\text{ps}}$ decreases with $\alpha$. Moreover, it is easy to check the condition (\ref{ccdc}).

\begin{figure}
\center{\subfigure[]{\label{PA0xy}
\includegraphics[width=5cm]{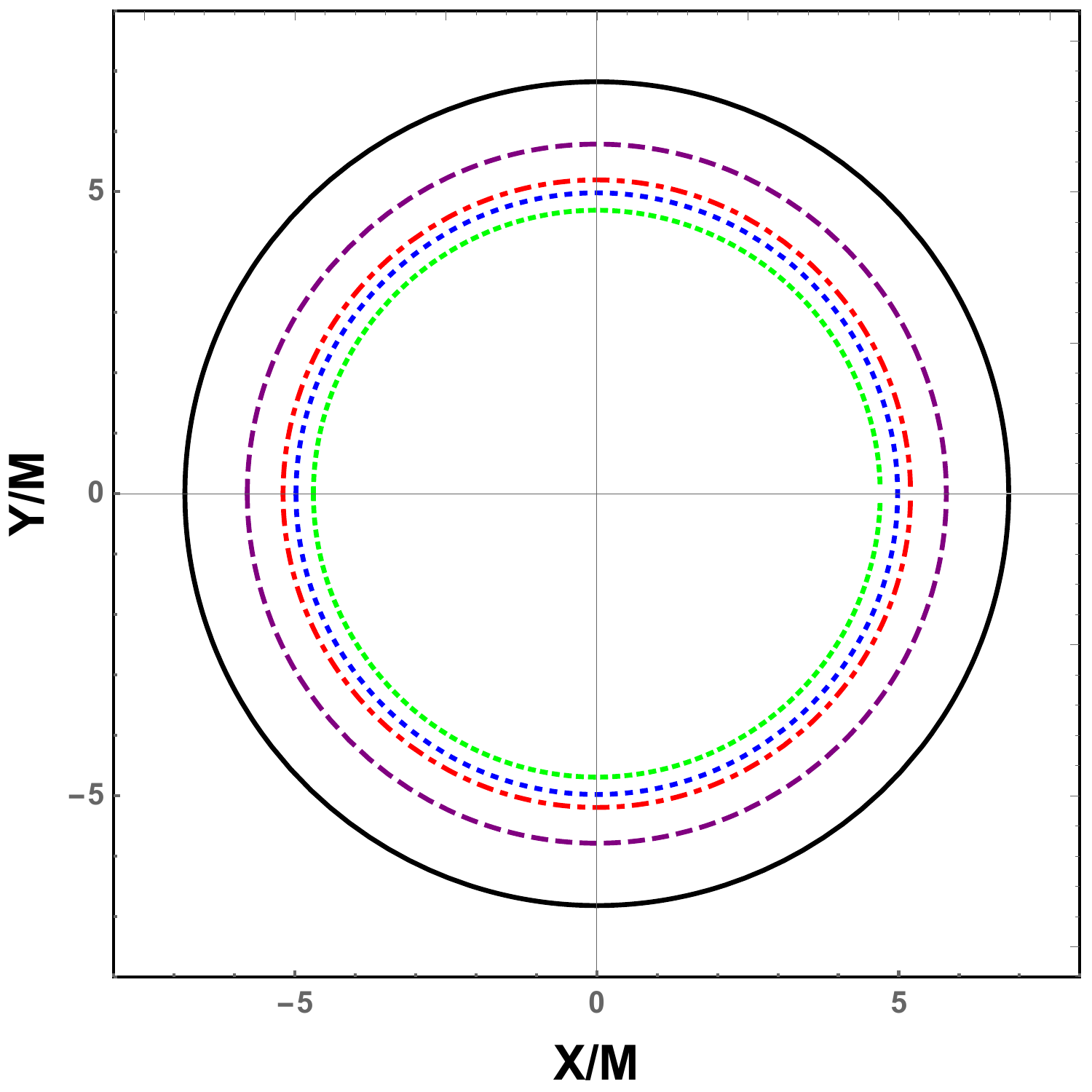}}
\subfigure[]{\label{PA0rsa}
\includegraphics[width=7cm]{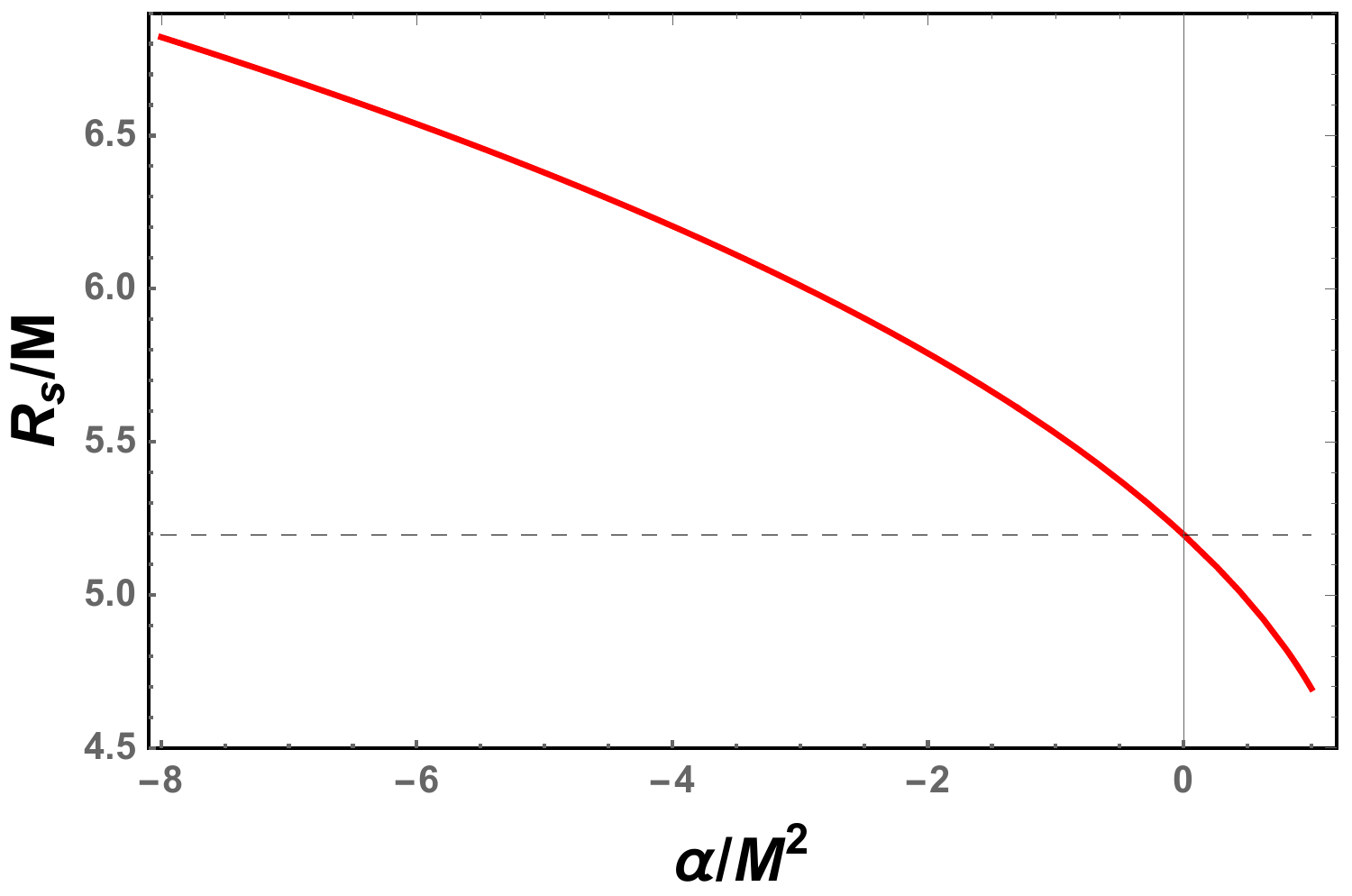}}}
\caption{(a) Shadows cast by the nonrotating GB black holes with $\alpha/M^2$=-8, -2, 0, 0.5, and 1 from outside to inside. (b) The radius of the shadows as a function of $\alpha$.}\label{ppA0rsa}
\end{figure}

The black hole shadow is round in this case. Employing the photon sphere radius $r_{\text{ps}}$, we can find the radius of the black hole shadow. After a simple calculation, we obtain the equation describing the boundary of the black hole shadow, which is given by
\begin{eqnarray}
 X^2+Y^2=R_s^2,
\end{eqnarray}
where $R_s$ denotes the radius of the shadow
\begin{eqnarray}
 R_s=\sqrt{\frac{2 \alpha  r_{\text{ps}}^2}{2 \alpha
   -\sqrt{8 \alpha
   r_{\text{ps}}+r_{\text{ps}}^4}+r_{\text{ps}
   }^2}}.
\end{eqnarray}
In Fig. \ref{PA0xy}, we show the shadows for the nonrotating CGB black holes with $\alpha/M^2$=-8, -2, 0, 0.5, and 1 from outside to inside. Obviously, the shadow shrinks with the increase of $\alpha$. The radius $R_s$ of the shadow is also plotted as a function of $\alpha$ in Fig. \ref{PA0rsa}. We can see that positive $\alpha$ shrinks the shadow and negative one enlarges the shadow.

In the following subsection, we will show that the influence of the black hole spin on the size of the shadow is very tiny, and the shadow size is mainly dependent on the metric parameter $\alpha$.

\subsection{Rotating black hole shadows}

When the black hole spin is included, the shape of the shadow behaves quite differently. The critical photons moving from two different sides of the black hole have different values of $\xi$ and $\eta$. This effect makes the black hole shadow be elongated in the direction of the spin axis. So for a rotating black hole shadow, its shape is not a round but a distorted one. In this subsection, we focus on the shadow shapes cast by the rotating CGB black holes.

\begin{figure}
\center{\subfigure[$\theta_0=\frac{\pi}{2}$, $a/M$=0.5]{\label{PAxy9005}
\includegraphics[width=7cm]{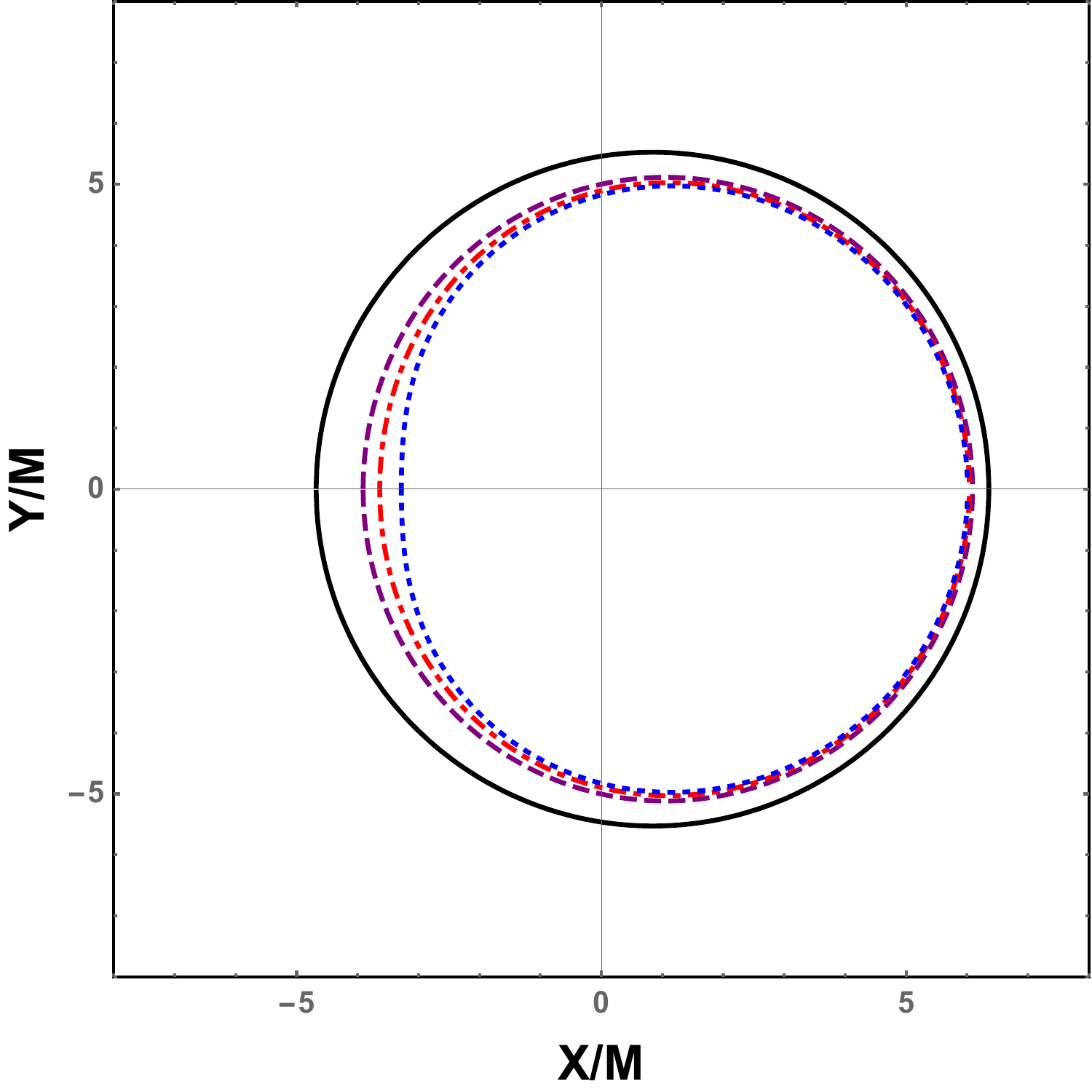}}
\subfigure[$\theta_0=\frac{\pi}{2}$, $a/M$=0.9]{\label{PAxy9009}
\includegraphics[width=7cm]{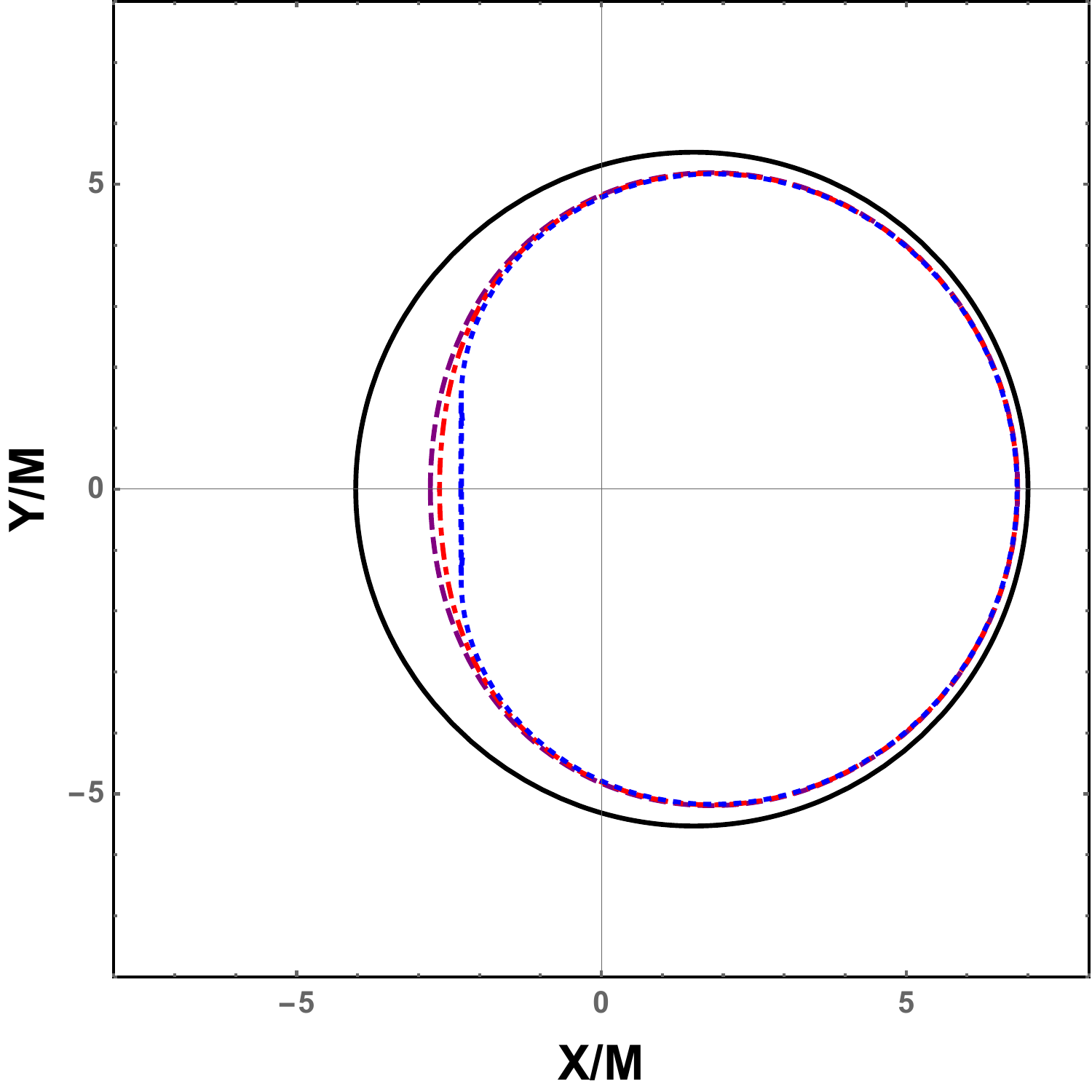}}
\subfigure[$\theta_0=\frac{\pi}{6}$, $a/M$=0.5]{\label{PAxy3005}
\includegraphics[width=7cm]{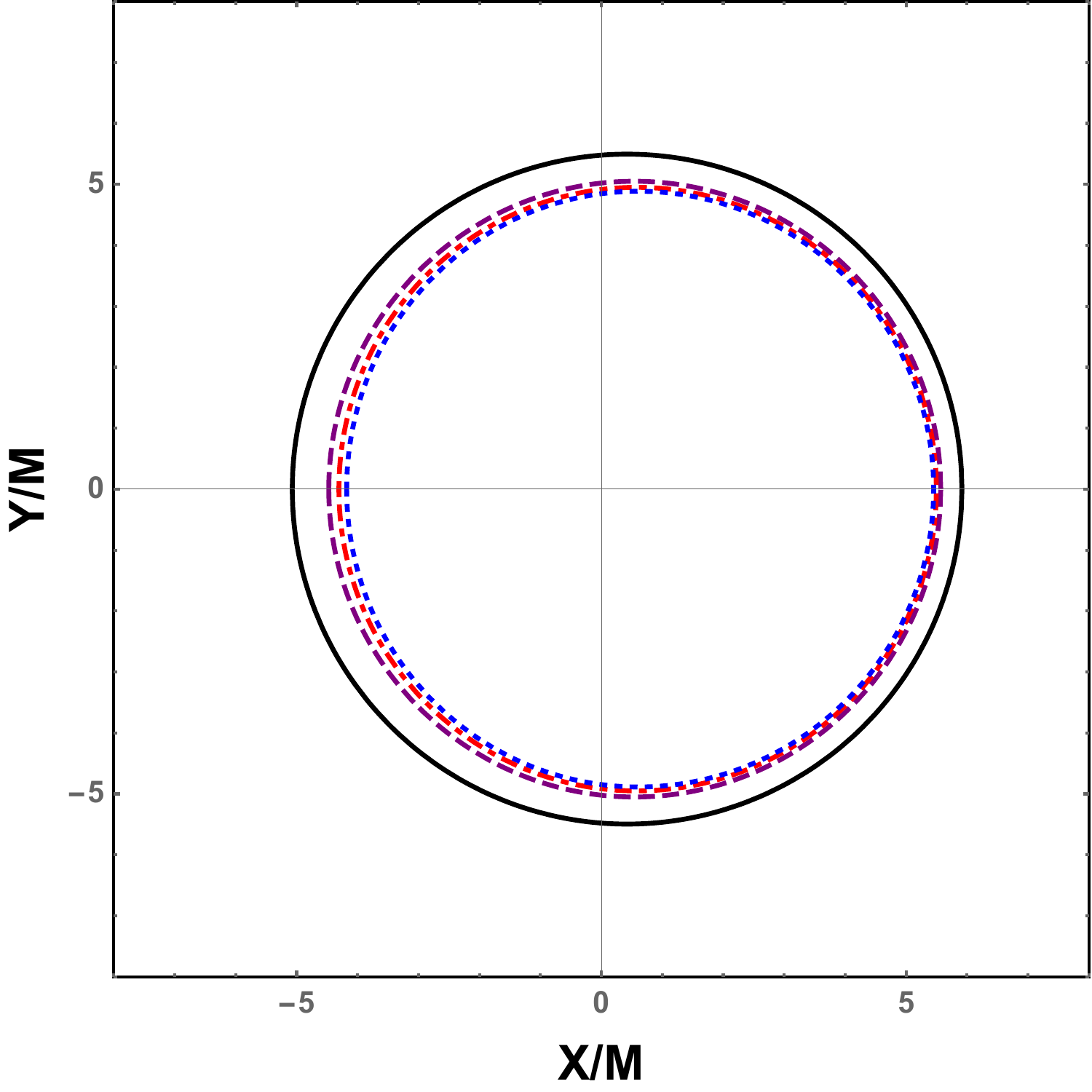}}
\subfigure[$\theta_0=\frac{\pi}{6}$, $a/M$=0.9]{\label{PAxy3009}
\includegraphics[width=7cm]{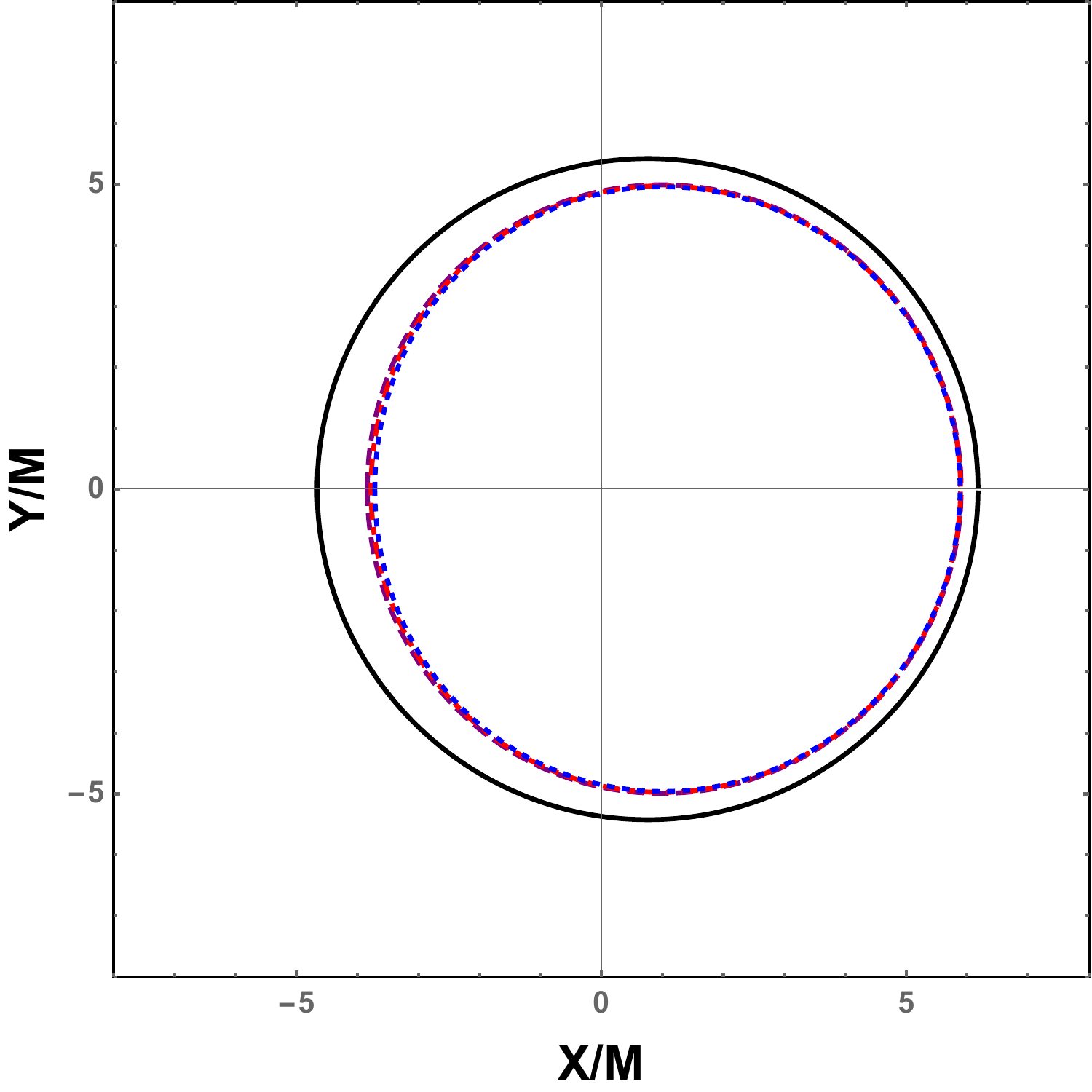}}}
\caption{Shadows of the rotating CGB black holes. In the left panel, $\alpha/M^2$=-1.0, 0.2, 0.4, and 0.5157 from outside to inside. In the right panel, $\alpha/M^2$=-1, 0.01, 0.04, and 0.0648 from outside to inside.}\label{ppAxy3009}
\end{figure}

Making use of (\ref{alpha}) and (\ref{beta}), we plot the shadows for the rotating CGB black holes in Fig. \ref{ppAxy3009}. All these figures show that the shadow shrinks with the increase of $\alpha$. Moreover, when the black hole spin approaches its maximal values, the shadows get more deformed. However, when the inclination angle of the observer decreases such that $\theta_0=\pi/6$, the shadows get less deformed when its spin approaches the maximal value. As a brief summary, we can conclude that the black hole shadow size is mainly dependent of $\alpha$, while its distortion is dependent of the black hole spin $a$.

\begin{figure}
\center{
\includegraphics[width=9cm]{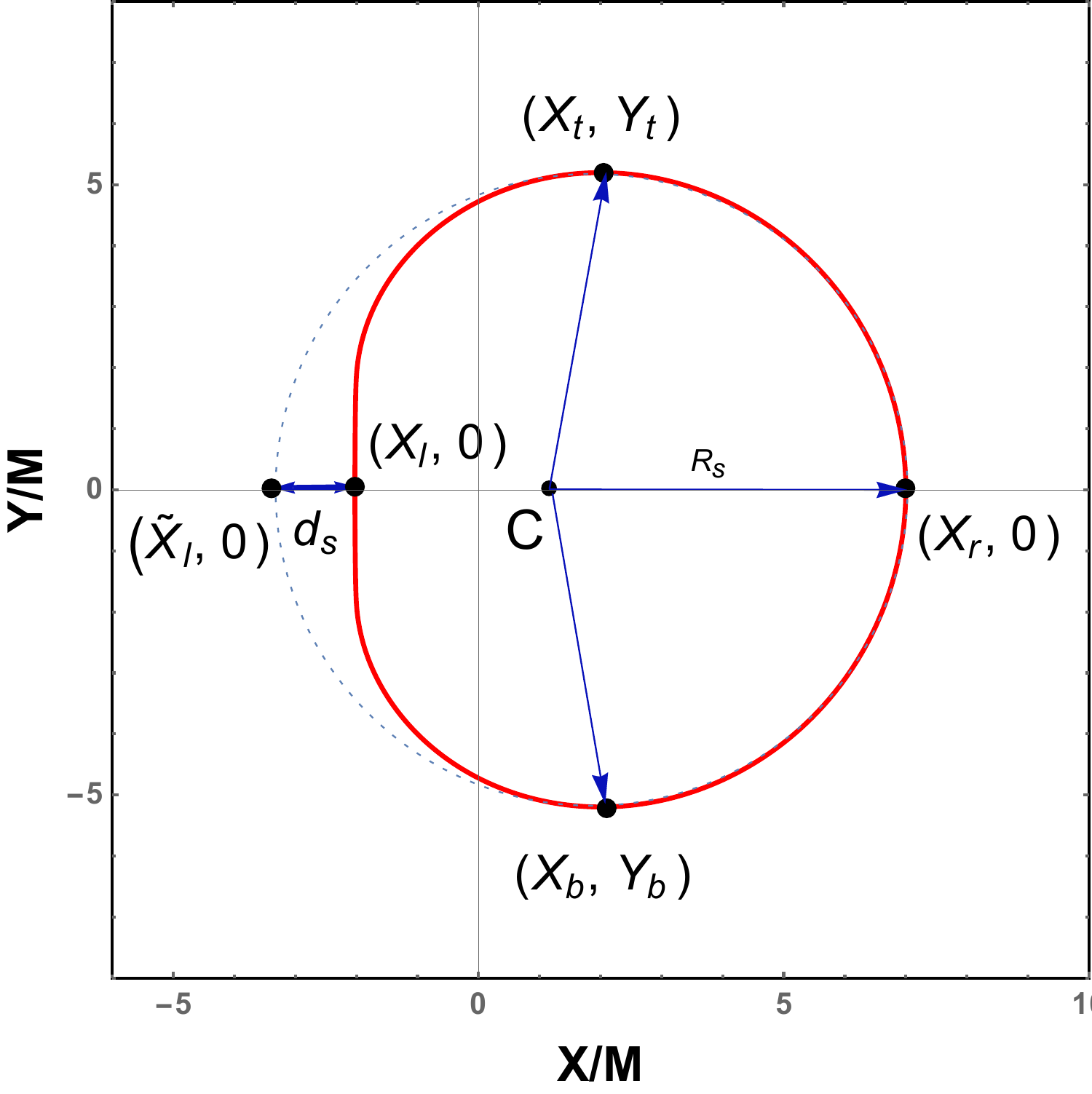}}
\caption{Schematic picture of the black hole shadow with nonvanishing spin.}\label{ppschematic}
\end{figure}

In order to get the black hole parameters through fitting the observed data, observables play a key role. The size and distortion are two important aspects of the shadows. So most of the observables are constructed through this fact. For clarity, the schematic picture of the black hole shadow is given in Fig. \ref{ppschematic} for nonvanishing spin. There are four characteristic points, the right point ($X_r$, 0), left point ($X_l$, 0), top point ($X_t$, $Y_t$) and bottom point ($X_b$, $Y_b$) of the shape. Considering the symmetry, one has $X_t=X_b$ and $Y_t=-Y_b$. As suggested in Ref. \cite{Hioki}, the size of the shadow can be approximately measured by the radius $R_s$ of the reference circle, which is just the circle passing the right, top, and bottom points of the shadow. Its center C is also on the $X$ axis. The reference circle cuts the $X$ axis at ($\tilde{X}_l$, 0). For a shadow cast by a nonrotating black hole, we will have $\tilde{X}_l=X_l$. However, the spin will distinguish these two points, and the shadow shape gets distorted. Another observable $\delta_s$ is also given in Ref. \cite{Hioki} to measure the distortion of the shadow with respect to the reference circle. In the following, we will study these observables as function of the metric parameter $\alpha$.

From the geometry of the shadow, the observables $R_s$ and $\delta_s$ can be expressed with the coordinates of these characteristic points as
\begin{eqnarray}
 R_s&=&\frac{(X_t-X_r)^2+Y_t^2}{2(X_r-X_t)},\label{rrrs}\\
 \delta_s&=&\frac{d_s}{R_s}=\frac{X_l-\tilde{X}_l}{R_s}.
\end{eqnarray}
Recently, the ratio $k$ of two diameters $\Delta X$ and $\Delta Y$ attracts much more attention. It is a new observable that can be fitted by the data of M87*. In terms of these coordinates, the ratio reads
\begin{equation}
 k_s=\frac{\Delta Y}{\Delta X}=\frac{Y_t-Y_b}{X_r-X_l}=\frac{2Y_t}{(2-\delta_s)R_s}.
\end{equation}
where we have used the relations $\tilde{X}_l=X_r-2R_s$. So it is clear that these observables are not independent of each other. As shown in Fig. \ref{ppAxy3009}, we can see that if a black hole spin is not very close to its maximum, the shadow is almost round. This conclusion is more true when the observer leaves the equatorial plane. Adopting this result, one has $R_s\approx2Y_t$, and thus $k_s\approx1/(2-\delta_s)$. Moreover, since $\Delta Y\geq\Delta X$, we have $k_s\geq1$.

\begin{figure}
\center{\subfigure[$\theta_0=\frac{\pi}{2}$]{\label{POds90}
\includegraphics[width=7cm]{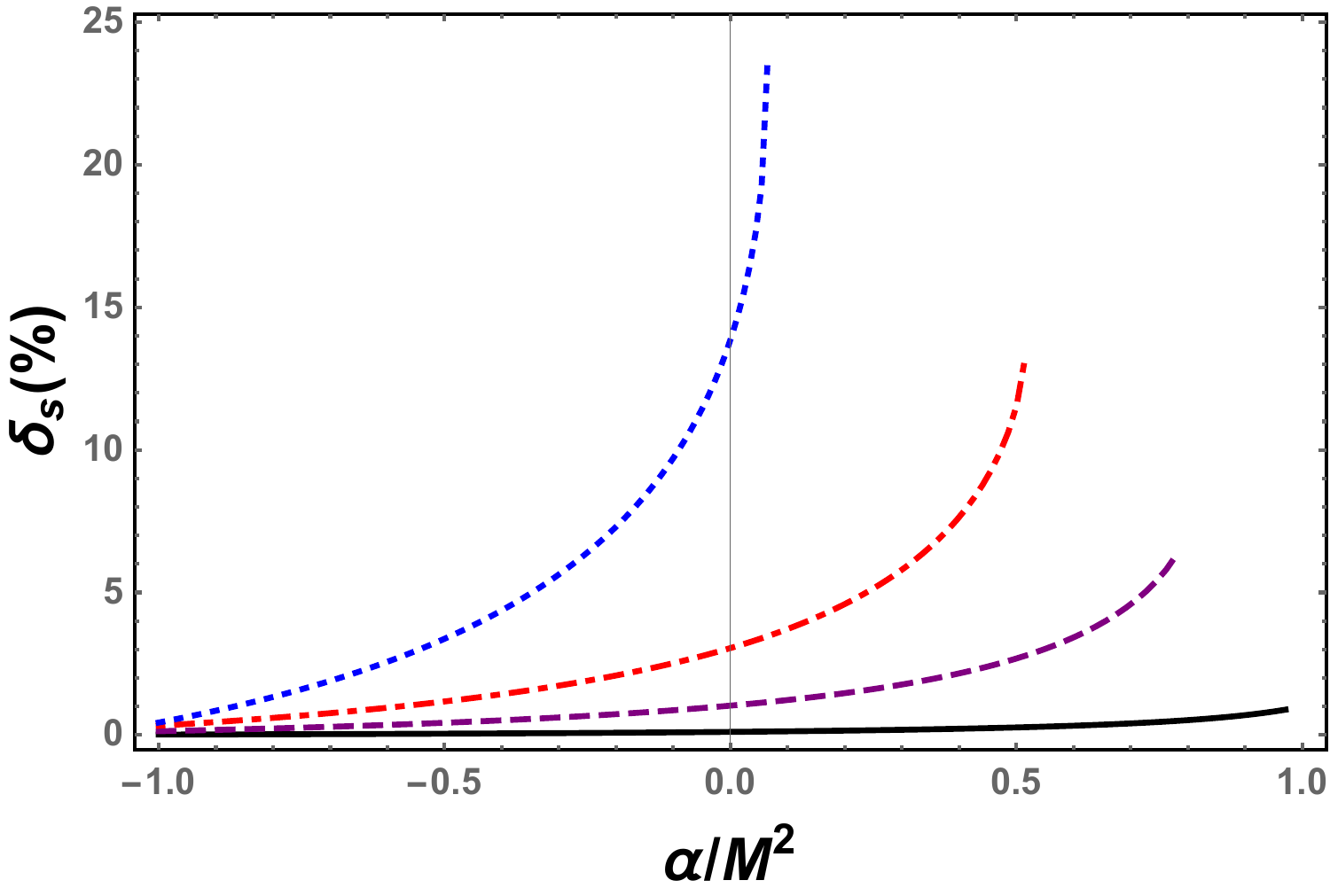}}
\subfigure[$\theta_0=\frac{\pi}{2}$]{\label{POks90}
\includegraphics[width=7cm]{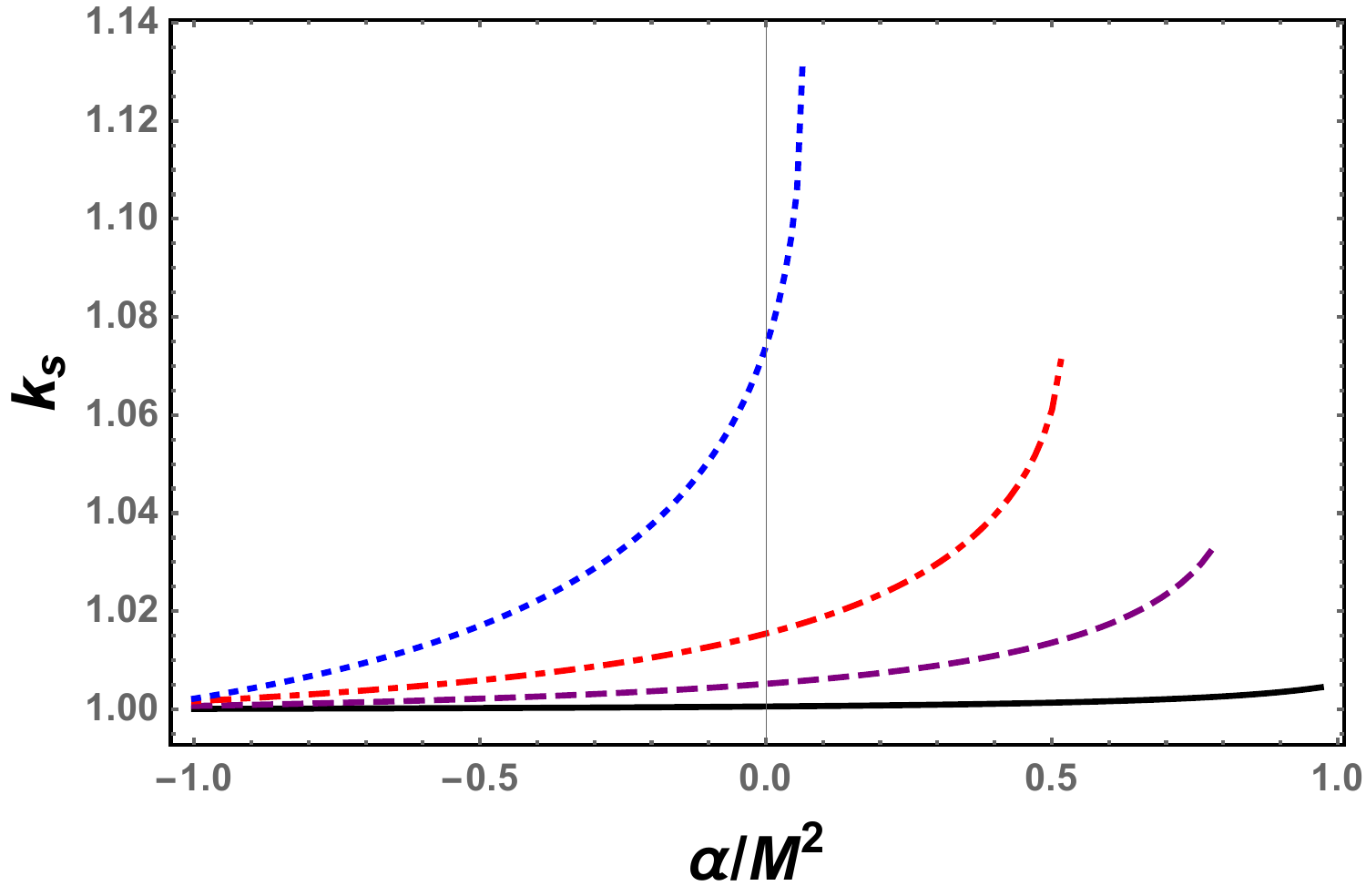}}
\subfigure[$\theta_0=\frac{\pi}{6}$]{\label{POds30}
\includegraphics[width=7cm]{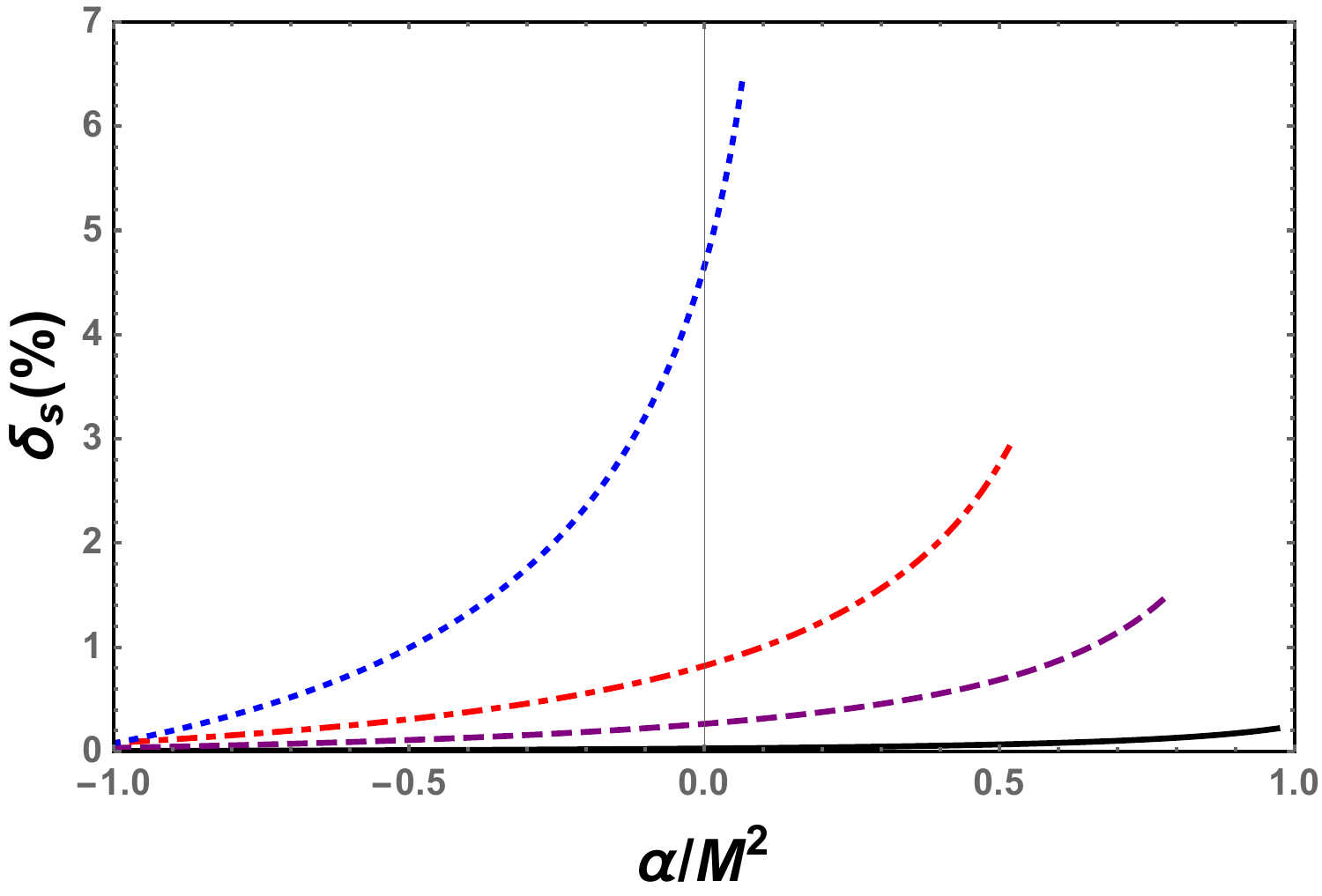}}
\subfigure[$\theta_0=\frac{\pi}{6}$]{\label{POks90}
\includegraphics[width=7cm]{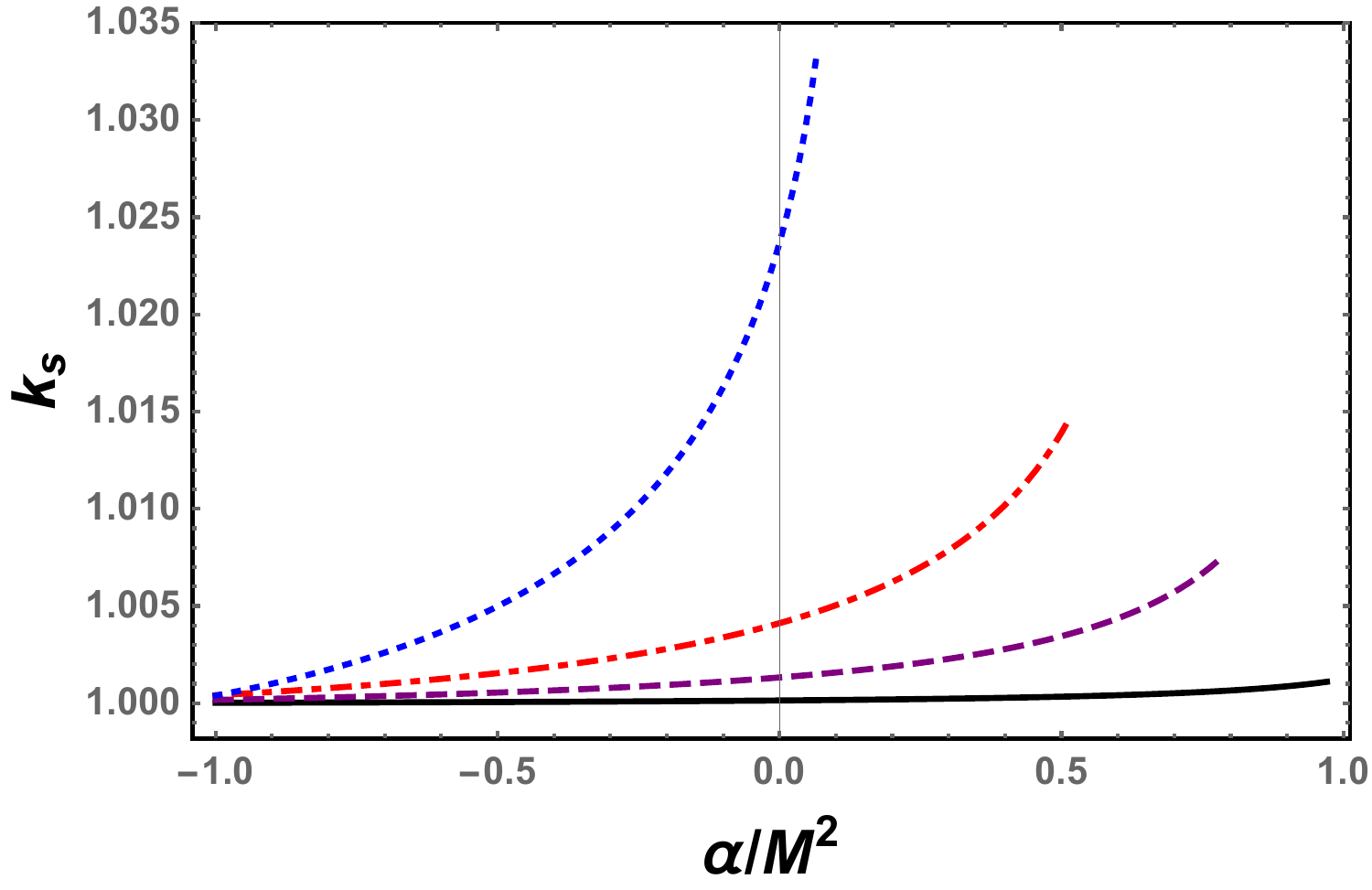}}}
\caption{Observables $\delta_s$ and $k_s$ for the rotating CGB black hole shadows. The spin is set as $a/M$=0.1 (black solid curve), 0.3 (dash purple curve), 0.5 (dotted dash red curve), and 0.9 (dotted blue curve) from bottom to top.}\label{ppOds90}
\end{figure}

For black hole spin $a/M$=0.1, 0.3, 0.5, and 0.9, we calculate these observables when $\theta_0$=$\pi/2$ and $\pi/6$. The results indicate that the influence of the spin is mainly on the maximum value of $\alpha$, while on the shadow size $R_s$ is very tiny. So we will not show them here. The behaviors of observables $\delta_s$ and $k_s$ are exhibited in Fig. \ref{ppOds90}. From the figures, we find that both $\delta_s$ and $k_s$ increase with $\alpha$ and $a$, while decrease with $\theta_0$. Further comparing with the Kerr black hole case, positive $\alpha$ will increase $\delta_s$ and $k_s$, while negative ones decrease them. For example, when $a/M$=0.9, the Kerr black hole has $\delta_s$=13.87\% and $k_s$=1.07 for $\theta_0=\pi/2$. Meanwhile, the rotating CGB black hole can achieve $\delta_s$=23.61\% and $k_s$=1.13, respectively. One can expect, these values get larger for high black hole spin.

In summary, we obtain the following result: 1) the shadow size $R_s$ is mainly dependent of the metric parameter $\alpha$. Positive $\alpha$ decreases the size, while negative one increases it. 2) When the black hole approaches its extremal case, the shadow will have larger values of $\delta_s$ and $k_s$. We believe these results will provide us the information on how to constrain the black hole spin $a$ and the parameter $\alpha$.

Before ending this section, we would like to note that here we only limit our attention in regions I and II, where the black hole always has one horizon at least. In region III, it denotes the naked singularity. Similarly, a naked singularity can also cast a shadow in the sky of an observer. However, its shadow is significantly different from the black hole shadow. Generally, the shadow shape of a naked singularity will not be a two-dimensional dark zone, while a one-dimensional opened dark arc. Considering that in a realistic scenario, the neighborhood of the arc will also be darkened, and thus there will be a dark lunate shape for the naked singularity. On the other side, if the value of the spin does not exceed the extreme case too much, the shape might similar to an extremal black hole. Surely, there is no evidence that the naked singularity can exist in our universe beyond the cosmic censorship conjecture. It is also worth to consider the shadow cast by a naked singularity. However, we only consider the black hole shadow in this paper, and the shadow of the naked singularity is a valuable issue for future.

\section{Shadow of M87* and Gauss-Bonnet coupling}
\label{M87}

In this section, we would like to use the result of the EHT Collaboration to fit the metric parameter $\alpha$ by the observation of M87*.

For M87*, its observed shadow has an angular diameter of 42 $\pm$3 $\mu$as. The observation of the jet indicates that the inclination angle is about 17$^\circ$ \cite{Walker}. Based on the stellar population measurements, the distance $D$ of M87* from us was estimated to be $D$=16.8$\pm$0.8 Mpc \cite{Blakeslee,Bird,Cantiello}. The stellar dynamics and gas dynamics studies showed that the mass of M87* is about $6.2^{+1.1}_{-0.5}\times10^9$ $M_{\odot}$ \cite{Gebhardt} and $3.5^{+0.9}_{-0.3}\times10^9$ $M_{\odot}$ \cite{Walsh}, respectively. Meanwhile, the EHT Collaboration reported that the mass of M87* is 6.5$\pm0.7\times10^9$ $M_{\odot}$. Their result also implies that the absolute value of the dimensionless black hole spin $a/M$ is in the range (0.5, 0.94).

For simplicity, we adopt the following data, $D$=16.8 Mpc, $M$=6.2$\times10^9$ $M_{\odot}$, and 6.5$\times10^9$ $M_{\odot}$. The inclination angle is chose to be 17$^\circ$. According to the Blandford-Znajek mechanism, the jet of M87* is powered by the black hole spin. Thus we suppose that the inclination angle equals to the jet angle, and this result holds for this four-dimensional GB black hole. Although as we show above, the rotating black hole solution only obeys the vacuum field equation on the equatorial plane. However, note that when matter fields are included, the field equation could be satisfied. So the choice of the inclination angle deviating from $\pi/2$ may be appropriate. Then we examine the average angular diameter $D_s=2R_s$ with $R_s$ given by (\ref{rrrs}) to fit the shadow size 42 $\pm$3 $\mu$as, or even 10\% offset is considered.

\begin{figure}
\center{\subfigure[]{\label{PFitaN7}
\includegraphics[width=7cm]{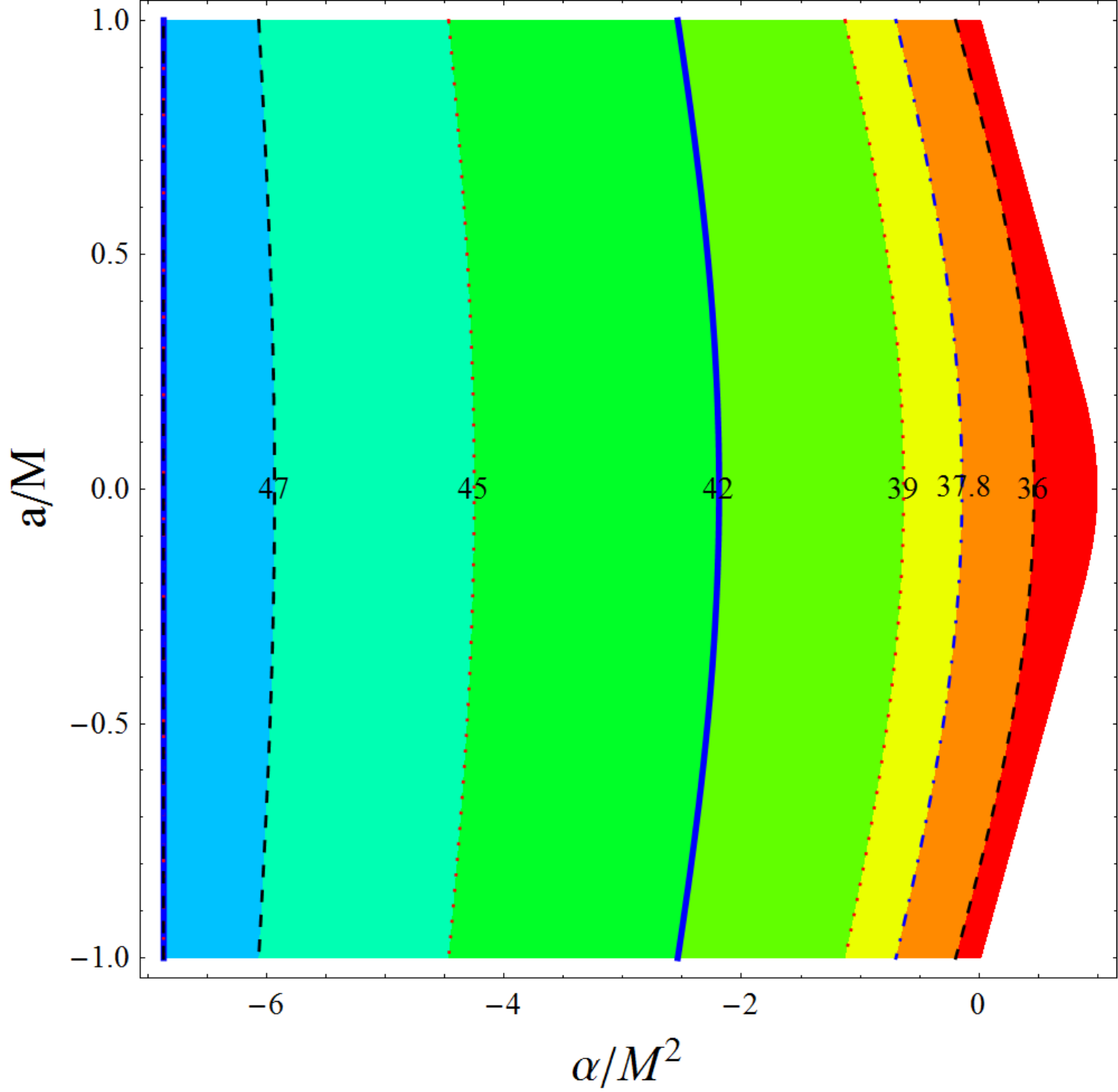}}
\subfigure[]{\label{PFitaP1a}
\includegraphics[width=7cm]{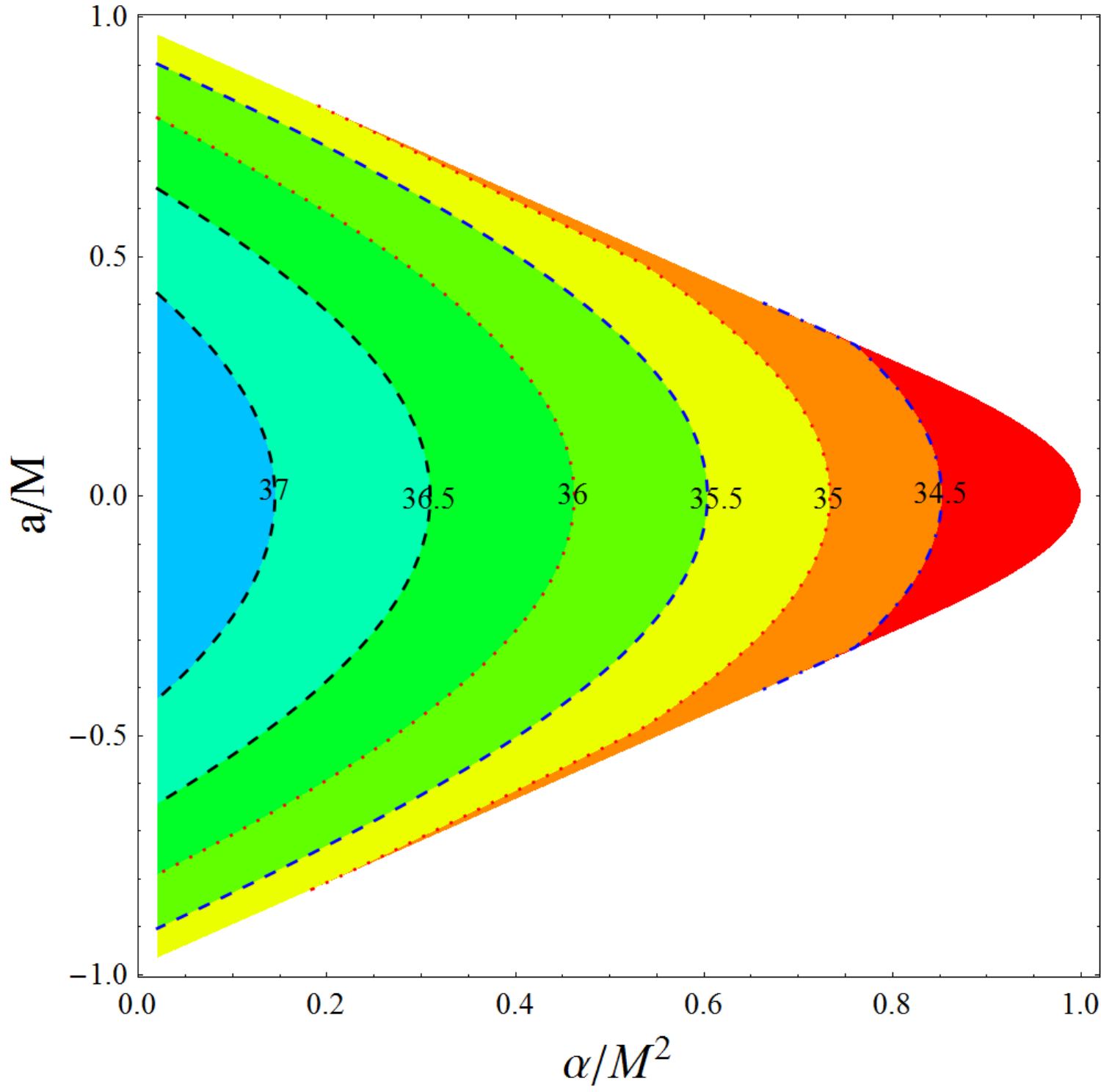}}}
\caption{Contours of the angular diameter of M87* in $a/M$-$\alpha/M^2$ plane with $D$=16.8 Mpc, $M=6.2\times10^9$ $M_{\odot}$. (a) $\alpha/M^2\in$ (-7, 1). (b) $\alpha/M^2\in$ (0, 1).}\label{ppFitaP1}
\end{figure}

First, we take $M$=6.2$\times10^9$ $M_{\odot}$ from the stellar dynamics, and show the contours of the angular diameter $D_s$ of M87* in $a/M$-$\alpha/M^2$ plane in Fig. \ref{ppFitaP1}. From Fig. \ref{PFitaN7}, we see that $\alpha/M^2$ is about in the range (-4.5, -1) for the observed shadow diameter 42 $\pm$3 $\mu$as. Moreover even if the 10\% offset of the diameter is taken into account, which reduces the corresponding angular diameter to 37.8 $\mu$as, the associated parameter $\alpha/M^2$=-0.28 and -0.62 for $|a|/M$=0.5 and 0.9, respectively. Therefore, the observation favors negative values of $\alpha$. For clarity, we also show the contours in $\alpha/M^2\in$ (0, 1) in Fig. \ref{PFitaP1a}. It is evident that in the positive range of $\alpha$, the diameter is about in 34$\sim$37.5$\mu$as, which obviously lies out the observation.

\begin{figure}
\center{\subfigure[]{\label{PFitaN7B}
\includegraphics[width=7cm]{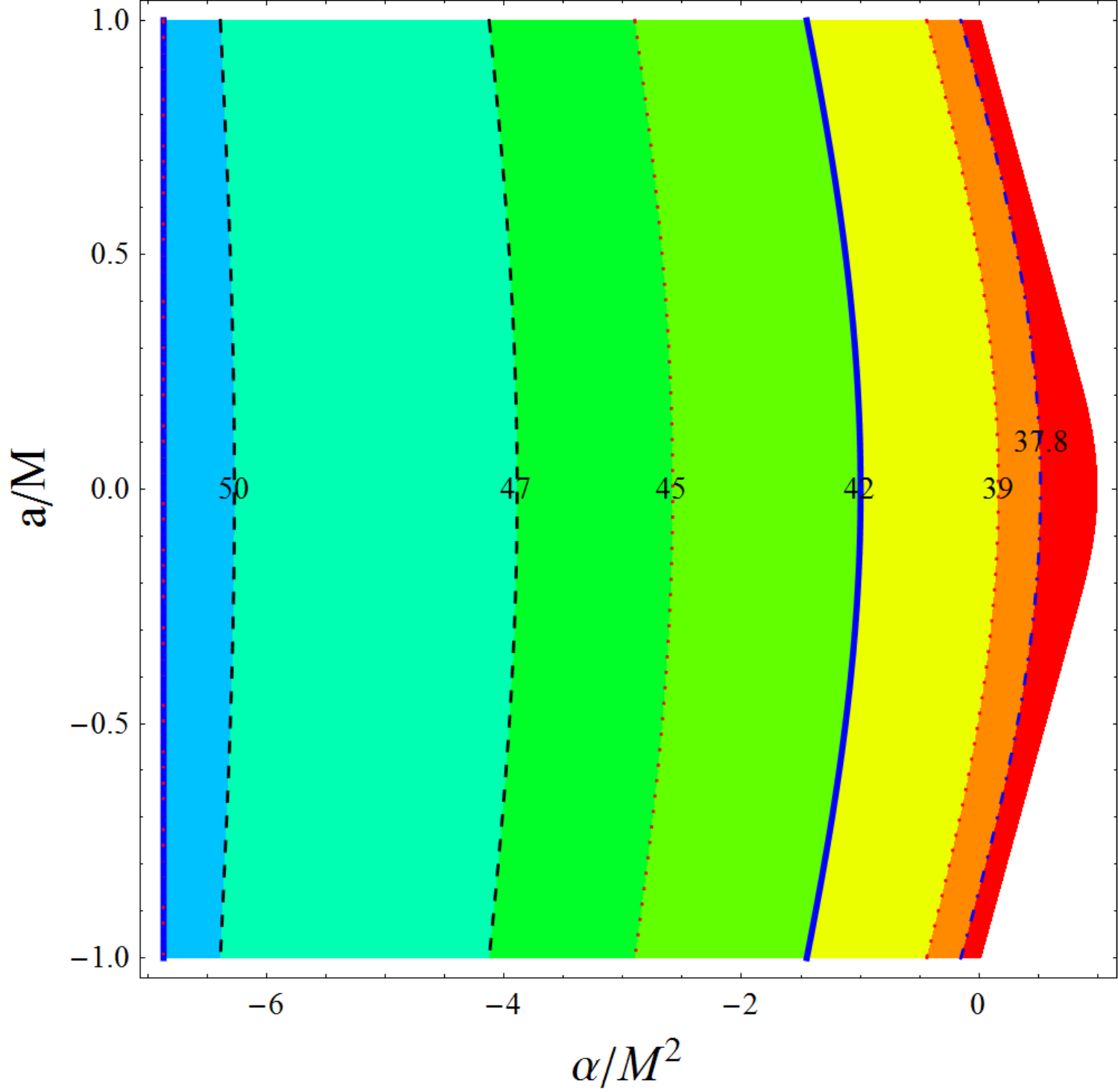}}
\subfigure[]{\label{PFitaP1aB}
\includegraphics[width=7cm]{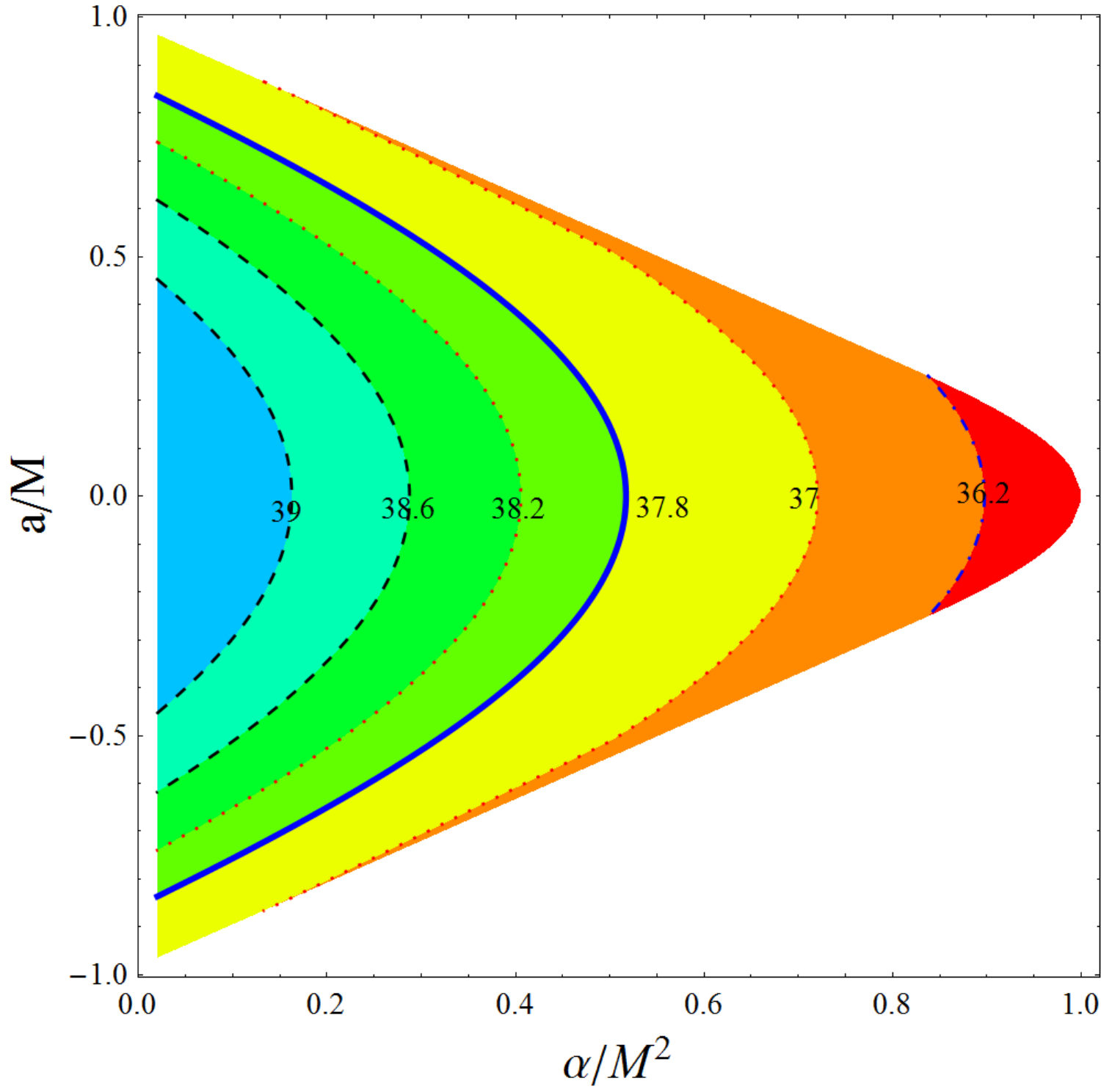}}}
\caption{Contours of the angular diameter of M87* in $a/M$-$\alpha/M^2$ plane with $D$=16.8 Mpc, $M=6.5\times10^9$ $M_{\odot}$. (a) $\alpha/M^2\in$ (-7, 1). (b) $\alpha/M^2\in$ (0, 1).}\label{ppFitaP1B}
\end{figure}

When taking the mass of M87* $M$=6.5$\times10^9$ $M_{\odot}$ given by the EHT Collaboration, we plot the corresponding contours in Fig. \ref{ppFitaP1B}. For $D_s\in(39, 45)\mu$as, it also falls in the range of negative $\alpha$. While taking into account of the 10\% offset of the diameter such that $D_s=37.8\mu$as, $\alpha/M^2$ takes 0.26 for $a/M$=0.5. However when $a/M>$0.83, $\alpha$ will become negative, i.e., $\alpha/M^2$=-0.05 for $a/M$=0.94.

In summary, combining with the mass estimated from the stellar dynamics or by the EHT Collaboration, the observation favors $\alpha/M^2\in$(-4.5, 0) when considering the black hole spin $a/M\in$(0.5, 0.94). Therefore, modeling M87* with a rotating CGB black hole, astronomical observation favors a negative $\alpha$.

\section{Conclusions and discussions}
\label{Conclusions}

In this paper, we first constructed a four-dimensional rotating CGB black holes by using the NJ algorithm. Then, we studied the geodesics of a particle moving in this background. Employing the null geodesics, we investigated the shadow cast by nonrotating and rotating black holes. At last, by combining with the observation of M87*, we constrained the possible range of the black hole parameter $\alpha$.

For the nonrotating black hole in the case of positive $\alpha$, the property of its horizon is similar to the charged black hole. Black hole with two horizons and naked singularity are bounded by the extremal black hole with $\alpha=M^2$. While for the negative $\alpha$ case, it was found that at the short radial distances, the spacetime has no real solution. However, if this range is hidden behind the horizon, the solution appears as a well behaved external solution. Thus, the study can be extended to $\alpha/M^2$=-8.

For the rotating black hole case, the horizon will be deformed by the black hole spin. However, the major property is still unchanged. As shown above, we have $-8\leq\alpha/M^2\leq1$ and $-2\leq a/M\leq2$. The particular details were given in Fig. \ref{ppPT}. As shown above, negative $\alpha$ indicates a negative energy, and thus such interesting implications extend the study on this modified gravity.

The geodesics of a particle moving in the CGB black hole background was solved following the Hamilton-Jacobi approach, which was also found to be a Kerr-like result. Based on the null geodesics, the shadow in the celestial coordinates was displayed. We observed that the shadow size is mainly dependent of $\alpha$, for example, positive $\alpha$ decreases the size while negative one increases it. On the other hand, the distortion mainly depends on the black hole spin and the extremal bound. The distortion $\delta_s$ and the ratio $k_s$ of these two diameters were also studied. Both them increase with $a$ and $\alpha$. Comparing with the Kerr black hole case, $\delta_s$ and $k_s$ increase with positive $\alpha$ and decrease with the negative one.

Finally, we modeled M87* with this rotating CGB black hole, and used the observation to constrain the possible range of the metric parameter $\alpha$. Considering the inclination angle $\theta_0=17^\circ$ from the
observation of the jet and the distance $D=16.8$Mpc from stellar population measurements, we plotted the contours of the angular diameter of the rotating CGB black hole with $M$=6.2$\times10^9$ $M_{\odot}$ and 6.5$\times10^9$ $M_{\odot}$ from the stellar dynamics and EHT Collaboration, respectively. Comparing with the angular diameter 42 $\pm$3 $\mu$as of M87*, our result supports that the parameter $\alpha$ is negative. Even if the 10\% offset of the diameter is taken into account, the most possible range still falls in the negative $\alpha$ region.

Although the EHT Collaboration molded M87* with the Kerr black hole, and confirmed that the observation supports the GR, it also leaves us a possible window to test modified gravity due to the resolution of the observation and the unknown mass of the accretion of M87*. Combing with these two different mass data of M87*, our results favor that the black hole parameter should fall in the possible negative range $\alpha/M^2\in$(-4.5, 0) or take very small positive values.

Note that the shadow was also studied in Ref. \cite{Kumar2} after us for the same black hole solution. They calculated the area and oblateness of the shadow and then found that the rotating black hole is consistent with M87* within a finite parameter space. Here we investigated the distortion and oblateness of the shadow, and then fitted the angular diameter with M87* data. It is worth to mention that in our approach, we also considered negative metric parameter $\alpha$. Although there is no real solution at short distance, it is always hidden behind the outer horizon. This also shows a potential understanding of the conformal anomaly gravity. Moreover, our study takes the first step towards the rotating black hole solution in the GB gravity.

\section*{Acknowledgements}

We would like to thank Dr. M. Guo for useful discussions about the rotating black hole. This work was supported by the National Natural Science Foundation of China (Grants No. 12075103, No. 11875151, and No 12047501), the 111 Project (Grant No. B20063), and the Fundamental Research Funds for the Central Universities (No. Lzujbky-2019-ct06).

\end{document}